\documentclass[lettersize,journal]{IEEEtran}
\usepackage{amsmath,amsfonts}
\usepackage{algorithmic}
\usepackage{algorithm}
\usepackage{array}
\usepackage[caption=false,font=normalsize,labelfont=sf,textfont=sf]{subfig}
\usepackage{textcomp}
\usepackage{stfloats}
\usepackage{url}
\usepackage{verbatim}
\usepackage{graphicx}
\usepackage{cite}
\usepackage{mathtools}
\usepackage{amssymb}
\usepackage{mathrsfs}
\usepackage{bbm}
\usepackage{color}
\usepackage{hyperref}
\usepackage{multirow}
\usepackage{booktabs} 

\hypersetup{
	colorlinks=true, 
	urlcolor=magenta,   
}

\newcommand{\reffig}[1]{Fig. \ref{#1}}
\newcommand{\reftable}[1]{Table \ref{#1}}
\newcommand{\refalg}[1]{Algorithm \ref{#1}}

\newcommand{\refsec}[1]{Section \ref{#1}}

\begin{document}
	
	\title{Effective Rank Analysis and Optimization of Flexible Antenna-Enabled Wireless Systems: Movable Antennas or Pinching Antennas?}
	\author{
		Cheng Yang 
		and Dong Li,~\IEEEmembership{Senior Member,~IEEE}
		\thanks{Cheng Yang and Dong Li are with the School of Computer Science and Engineering, Macau University of Science and Technology, Macau, China (e-mail: 3250010519@student.must.edu.mo; dli@must.edu.mo).}
		}
%
	
	\maketitle
	
	\begin{abstract}
        Flexible antenna technology has recently emerged as a key enabler for next-generation wireless communications, which can effectively exploit the spatial degrees of freedom (DoF). 
        However, existing conventional metrics (e.g., spectral and energy efficiency) cannot directly measure the variability for different flexible antenna structures for the spatial DoF.
        To objectively analyze and compare the spatial DoF of different flexible antennas, the effective rank is introduced as a metric for two major flexible antenna technologies, i.e., movable antenna (MA) and pinching antenna (PA)-enabled wireless systems, optimizing their antenna positions over multi-time slots.
        However, the inherent non-convexity and high computational complexity of the resulting effective rank maximization problems in MA and PA systems render them hard to solve.
        To circumvent these problems, we propose the graph attention implicit quantile network (GAIQN) and multi-agent graph attention Q-network (MAGAQN) algorithms to enhance the effective rank of MA and PA systems through antenna positioning, respectively.
        Meanwhile, the novel top-k action selection methods are designed to ensure collision-free between multiple MAs on the two-dimensional array plane or multiple PAs on the same waveguide.
        Simulation results validate the effectiveness and advancement of our proposed GAIQN and MAGAQN algorithms compared with benchmarks, which enhance the effective rank by at least 1.6\% and 1.3\%, respectively, while consistently ensuring collision-free between flexible antennas.
        Besides, under the same number of flexible antennas, the MA system supports a higher effective rank than the PA system, whereas the PA system offers greater stability in terms of achievable spatial DoF.

	\end{abstract}
	

	\section{Introduction}
		\IEEEPARstart{T}{he} expected success of heterogeneous services and related applications in future wireless networks, such as autonomous driving, holographic telepresence, and the metaverse, requires unprecedented massive user access, heterogeneous data traffic, high bandwidth efficiency, and low-latency services in 6G and beyond (B6G). 
		To achieve these goals, multiple-input multiple-output (MIMO) has evolved as an effective technology for enhancing the throughput, spectral efficiency, and reliability of wireless communication systems \cite{11267231}.
		Nevertheless, current MIMO systems mostly consist of the fixed-position antennas (FPAs), whose positions are set once deployed.
        The lack of adaptation to dynamic user distributions, blockage, and reflection variation inherently constrains the systems' ability to fully exploit the spatial degrees of freedom (DoF) within its coverage areas \cite{11142311}.

        Recently,  flexible antenna technologies, which enable radiating elements to achieve dynamic channel control through physical antenna repositioning, have emerged as a key advancement in the evolution of MIMO.
        Unlike traditional FPAs, typical flexible antenna systems, e.g., movable antennas (MAs) and pinching antennas (PAs), can be dynamically adjusted in the three-dimensional (3D) space to achieve flexible control over the positions or other characteristics of the flexible antenna \cite{11143302}.
   		Existing studies have progressively explored different flexible antenna systems and optimization objectives. 
        Specifically, MA systems can enable the local movement of antennas within a specified area, thereby reshaping wireless channels \cite{11048972}.
		In a multi-user uplink communication scenario \cite{11226954}, a cross-linked movable antenna architecture was proposed to enable collective antenna movement, thereby addressing the prohibitive hardware cost of conventional independently driven MA systems and significantly reducing the total transmit power compared with FPA schemes.
		A two-timescale optimization framework was developed in \cite{11164786} for the MA-enhanced multi-user MIMO downlink system, where instantaneous channel state information (CSI) at the receiver and statistical CSI at the transmitter were exploited for receive and transmit antenna position design, respectively, to maximize the average achievable sum rate.
		More recently, an energy-efficient design of MA-aided multi-user downlink communication systems had been investigated in \cite{11218873}, where transmit beamforming and antenna positions were jointly optimized under practical movement and power constraints to maximize energy efficiency.
	    Different from MA systems, PA systems employ a dielectric waveguide as the transmission medium, which simply adds separate dielectric materials to dynamically activate specific points along a waveguide \cite{11222687}.
        Considering both minimum mean square error (MMSE) decoding and successive interference cancellation (SIC), a fractional programming-based approach was proposed to enhance the uplink sum-rate by optimizing  PA positions for PA-assisted multi-user multiple-input
        single-output (MISO) uplink communications \cite{11175694}.
	    Meanwhile, a throughput maximization framework for multi-user PA systems was proposed in \cite{11184829}, where a joint optimization strategy for antenna positioning and quantity selection is designed by exploiting waveguide-induced phase shifts for coherent signal superposition without conventional phase shifters.
        Then, a multi-waveguide PA system was investigated for multi-user communications \cite{11300296}, where waveguide multiplexing, division, and switching transmission structures with joint beamforming design were proposed to solve max–min fairness problems, achieving significant improvements in the minimum achievable rate compared with conventional FPA systems.
	    
        Despite the performance gains offered by MA and PA systems, existing hardware implementations mainly rely on mechanical movements or switching mechanisms, which introduce the non-negligible  antenna repositioning delay and may degrade the real-time performance of wireless systems.
	    To mitigate the latency caused by mechanical antenna repositioning, electrically activated antenna structures have been recently studied. 
	    In particular, to overcome the limitations of mechanically reconfigurable MAs, the authors investigated reconfigurable pixel antenna (RPA)-based electronic MAs for multi-user communications, proposed partially-connected and fully-connected RPA-based electronic MA array structures, and developed a two-step multi-user beamforming and antenna selection scheme \cite{11095802}, with theoretical analysis revealing a negligible performance loss compared to mechanical MAs.
		A practical multi-waveguide PA system with discretely deployed PAs was investigated in \cite{11165763}, where joint waveguide assignment, antenna activation, power allocation and SIC decoding order were optimized through a game-theoretic framework to improve the system throughput.
        
        It should be noted that the performance of the aforementioned MA and PA systems is still predominantly evaluated in terms of conventional metrics, e.g., spectral and energy efficiency.
		These are inherently influenced not only by the spatial reconfigurability of antenna structures, but also by transmit power control and beamforming strategies, which makes it difficult to disentangle the performance gains brought by the antenna physical reconfiguration itself from those introduced by optimization algorithms or scheduling policies.
        Hence, the effective rank is adopted as a structure-oriented metric \cite{roy2007effective}, which not only decouples these effects but also reflects channel balance and the suitability of linear precoding schemes such as zero forcing (ZF) and MMSE.
        Meanwhile, there have been few studies investigating the effective rank of wireless systems.
		In \cite{10217398}, the authors investigated the effective rank maximization of MIMO channels via transmit antenna position optimization, where the electromagnetic channel including mutual coupling was modeled using the method of moments and the optimal antenna distributions were obtained through a genetic algorithm.
		An effective rank enhancement of reconfigurable intelligent surface (RIS)-assisted MIMO channels was investigated in \cite{10314137}, where an efficient RIS phase optimization scheme was proposed and experimentally validated via the ${2 \times 2}$ RIS-assisted MIMO prototype, demonstrating significant gains in both effective rank and achievable rate.
		Then, multiple active reconfigurable intelligent surfaces (ARISs) were exploited in \cite{11225901} to mitigate rank deficiency in strong line-of-sight (LoS) channels by maximizing the effective rank, where joint optimization of ARIS phase shifts and deployment positions was developed under both continuous and discrete phase-shift models, achieving near-optimal spectral efficiency.
        
        Nevertheless, the aforementioned works are restricted to the FPA, which has limited capability in adjusting the antenna position to exploit the spatial variations of wireless channels for performance enhancement. 
        Besides, there are currently no design guidelines on how to deploy MA or PA for practical flexible antenna-enabled wireless communications systems. 
        To fill the above gaps, in this paper, we investigate the effective rank to analyze flexible antenna systems over multi-time slots, and provide an effective rank-based comparison between MA and PA systems. 
        We also explore the intrinsic differences between these flexible antenna structures in shaping the channel eigenstructure and utilizing the spatial DoF.
		The main contributions of our work are summarized as follows:
		\begin{enumerate}
			\item We consider the electrically activated MA and PA-enabled wireless systems in a multi-user downlink communication scenario. 
            Specifically, to ensure a fair comparison and eliminate the antenna repositioning delay caused by mechanical movement, we investigate an electrically activated approach for both MA and PA systems.
            The objective is to exploit their respective spatial DoFs to improve overall wireless system performance.

			\item To enhance wireless system performance exclusively through the spatial DoF, this work introduces the effective rank as a metric for MA and PA-enabled wireless systems, while explicitly incorporating collision-free between multiple MAs on the two-dimensional (2D) array plane and multiple PAs on the same waveguide as constraints for their respective optimization models.
            By enforcing these constraints to eliminate the impact of heterogeneous collision probabilities arising from different flexible antenna structures, the effective rank-based formulations enable a fair comparison.
			
			\item Traditional optimization algorithms struggle to effectively handle the dynamics of the system over multi-time slots, which makes them unsuitable for solving the formulated effective rank maximization problems.
			Hence, we propose the graph reinforcement learning (GRL)-based MA dynamic positioning and multi-agent graph reinforcement learning (MAGRL)-based PA dynamic positioning schemes.
			Specifically, the graph attention implicit quantile network (GAIQN) and multi-agent graph attention Q-network (MAGAQN) algorithms are proposed to improve the effective rank for MA and PA systems, respectively.
            Meanwhile, the top-k action selection methods are developed to ensure strictly collision-free between multiple flexible antennas.
            
	\end{enumerate}

	The rest of the structure of this paper is organized as follows.
	\refsec{sec_system_model} presents the models and problem formulation.
	\refsec{sec_gaiqn} and \refsec{sec_magaqn}  introduce the proposed GRL-based MA dynamic positioning and MAGRL-based PA dynamic positioning schemes, respectively.
     \refsec{sec_simulation} details the simulation results. 
     \refsec{sec_conclusion} concludes this paper.
    
    \textit{Notations:}
    ${\left|  \cdot  \right|}$ and ${ \cdot || \cdot }$ denote the determinant and vector concatenation operation.
    ${\left\| {\cdot} \right\|_1}$ and ${\left\| {\cdot} \right\|_2}$ are the 1-norm and 2-norm operations, respectively.
    ${\left[{\cdot}\right]^{\top}}$ is a transpose operation.
    ${\mathbb{E}\left[{\cdot}\right]}$ represents the expectation operation.
    ${\lor}$ is the OR operation.
    ${\odot }$ is the Hadamard product.
    The operator ${Y \overset{D}{=} X}$ indicates that the random variable ${Y}$ is distributed according to the same law as ${X}$. ${\mathbb{I}{\left\{{u < 0}\right\}}}$ is an indicator function. Let ${\rm{MLP}}$ denote the fully connected layer.

	\section{System Model \label{sec_system_model}}
\begin{figure*}[!t]
		\centering
		\subfloat[]{\includegraphics[width=0.425\linewidth]{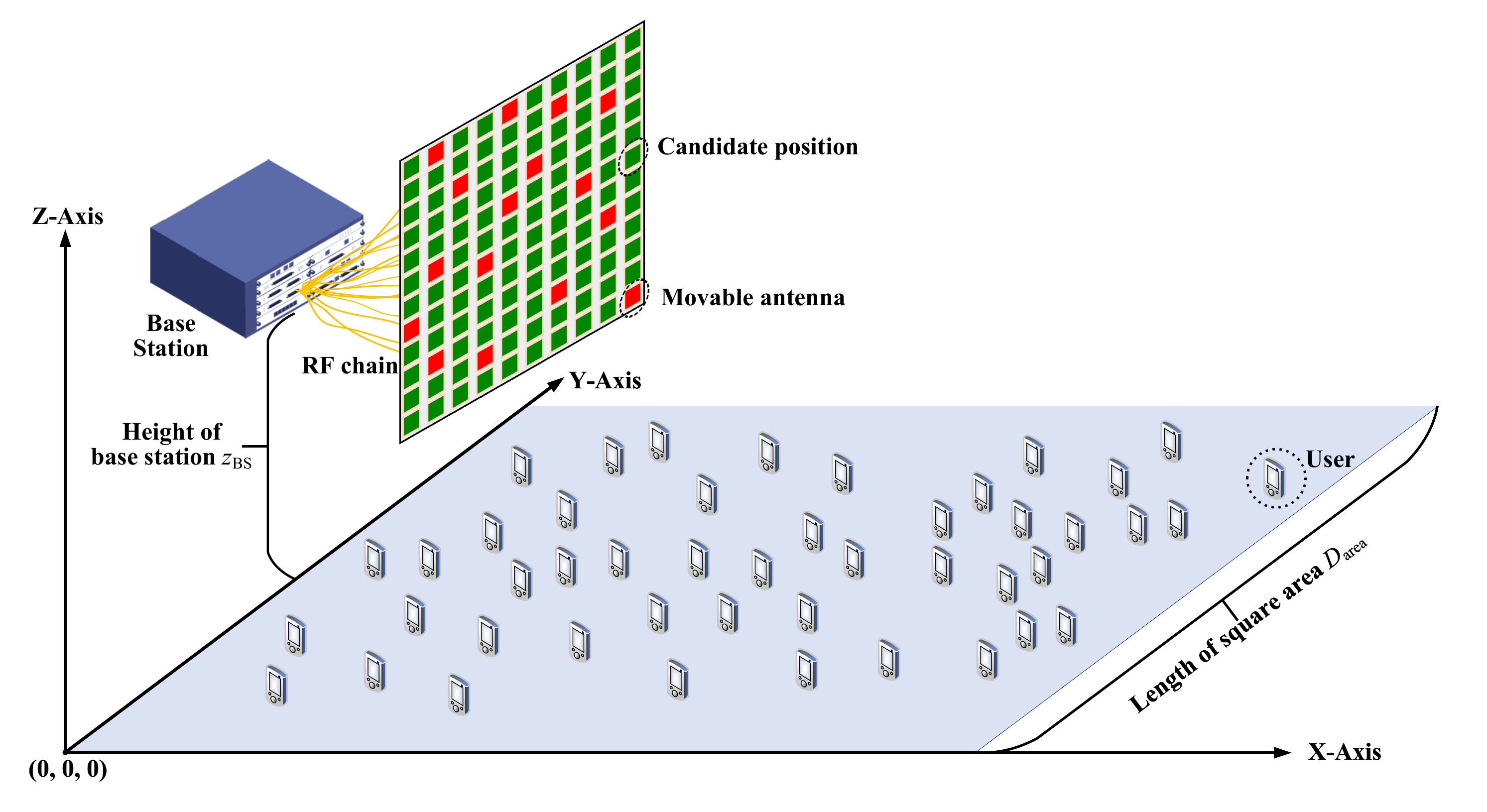}%
		\label{fig_system_model4mass}}
		\hfil
		\subfloat[]{\includegraphics[width=0.425\linewidth]{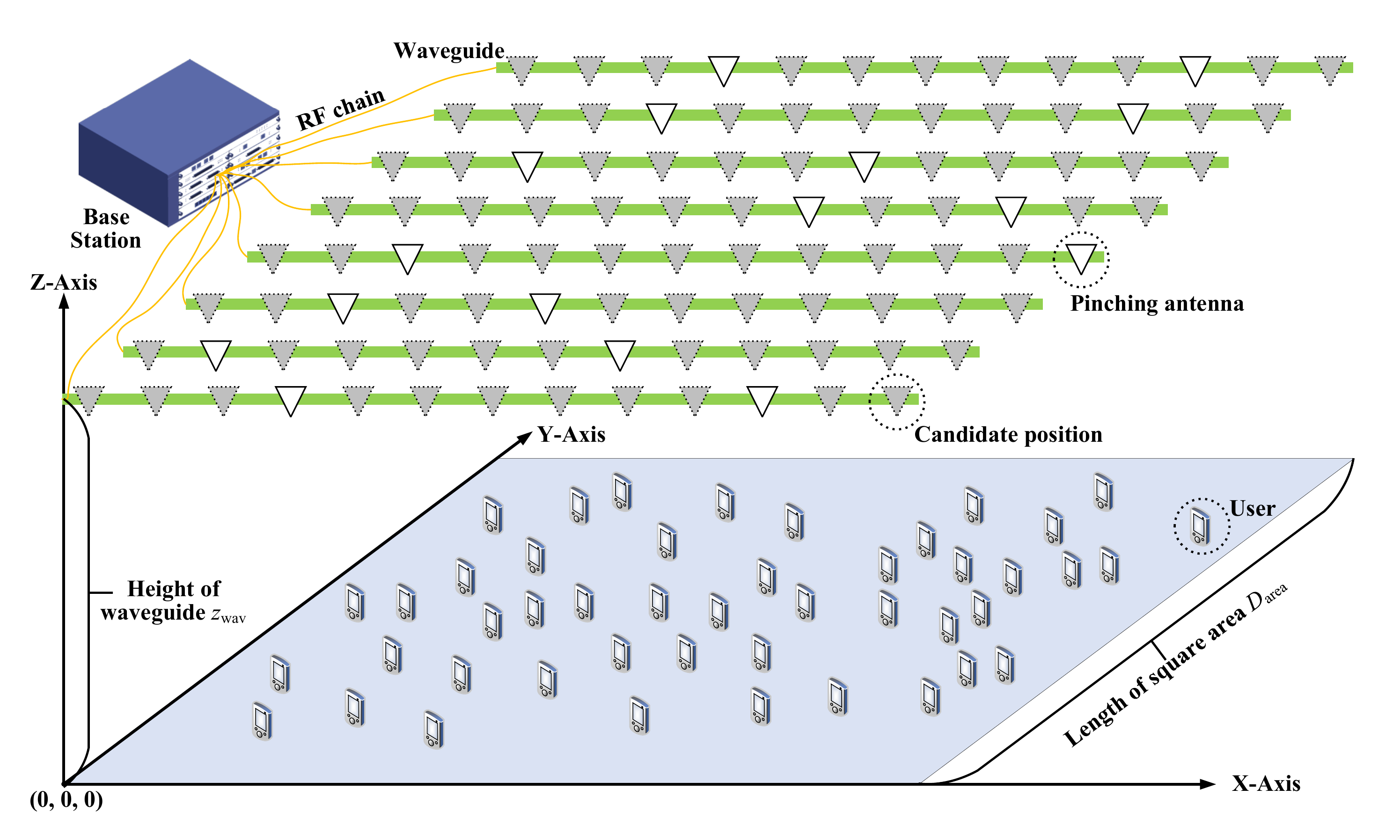}%
		\label{fig_system_model4pass}}
		\caption{Illustration of the considered flexible antenna-enabled wireless systems. (a) MA. (b) PA.}
		\label{fig_system_model}
		\end{figure*}

		As illustrated in \reffig{fig_system_model}, we consider a multi-user downlink communication scenario, where a base station (BS) equipped with an MA or a PA system serves ${N}$ single FPA users.
		To ensure fairness in comparing flexible antenna systems, the BS is assumed to be located in the center of the left side of the entire square area, where the length of the square area is ${D_{\text{area}}}$ m.
		Let ${\left[ {{x_{{\rm{BS}}}},{y_{{\rm{BS}}}},{z_{{\rm{BS}}}}} \right]^{\top}}$ denote the 3D position of the BS, and that of the ${n}$-th user is denoted as ${\boldsymbol{u}_{n}= \left[ {{x_n},{y_n},{z_n}} \right]^{\top} \in \mathbb{R}^{3 \times 1}}$ (${1 \le n \le {N}}$), where ${z_{\text{BS}}}$ and ${z_n}$ denote the antenna height of the BS and ${n}$-th user, respectively.
		
		\subsection{MA-enabled wireless system \label{sec_channel4ma}}
			In \reffig{fig_system_model}(a), the BS is equipped with a  2D array plane with ${M_{\text{ma}}}$ MAs.
			This work mainly considers an electrically activated approach to guarantee the low latency of antenna repositioning, whose 2D array plane can support ${I_{\text{r}}}$ rows and ${I_{\text{c}}}$ columns of candidate positions, i.e., all positions are constrained to a finite set \cite{11095802, 11048972}.
			Let ${\mathcal{P}_{\text{ma}} = \{ {{\bf{p}}_i^{{\rm{ma}}}, \cdots ,{\bf{p}}_{{I_{{\rm{pos}}}}}^{{\rm{ma}}}} \}}$ denote the set of candidate positions on the 2D array plane, where ${\forall {\bf{p}}_i^{{\rm{ma}}} \in \mathbb{R}^{2\times 1}}$, and ${I_{\text{pos}} = I_{\text{r}} \times I_{\text{c}}}$ indicates the total number of  candidate positions.
			Furthermore, to avoid coupling effects between MAs, the minimum row and column distance for all candidate positions is usually set as ${D_{\min}^{\text{ma}}=\lambda/2}$, where ${\lambda}$ denotes the carrier wavelength. 
			
			Consequently, it is established by the classical Cartesian coordinate system for MAs and users.
            For the ${m}$-th MA on the 2D array plane, its 2D position can be expressed by ${\boldsymbol{t}_m=\left[ {{x_m^{\text{ma}}},{y_m^{\text{ma}}}} \right]^{\top} \in \mathcal{P}_{\text{ma}}}$ (${1 \leq m \leq M_{\text{ma}}}$).
			Assuming that the far-field condition is satisfied for all MAs at different positions \cite{11048972}, the angle of departure (AoD) and the angle of arrival (AoA) remain unchanged for each channel path component.
			Let ${L^{\text{r}}}$ and ${L^{\text{t}}}$ denote the number of  receive and transmit channel paths from the BS to each user, respectively.

			In the ${{\ell}^{\text{t}}}$-th transmit channel path, its 2D transmit wavevector	${\boldsymbol{\kappa}_{n, {\ell}^{\text{t}}}^{\text{t}} \in \mathbb{R}^{2 \times 1}}$ from the BS to the ${n}$-th user is thus represented as
			\begin{equation}
				\boldsymbol{\kappa}_{n, {\ell}^{\text{t}}}^{\text{t}}=\frac{{2\pi }}{\lambda } \left[ {\cos {\theta ^{\text{t}}_{n,{{\ell}^{\text{t}}}}}\cos {\phi ^{\text{t}}_{n,{{\ell}^{\rm{t}}}}},\cos {\theta^{\text{t}} _{n,{{\ell}^{\text{t}}}}} \sin {\phi ^{\text{t}}_{n,{{\ell}^{\text{t}}}}}} \right]^{\top},
			\end{equation}
			where ${ 1 \le {\ell}^{\rm{t}} \le L^{\text{t}}}$. 
			${\theta^{\text{t}} _{n,{{\ell}^{\rm{t}}}}}$ and ${\phi^{\text{t}} _{n,{{\ell}^{\rm{t}}}}}$ are the elevation and azimuth AoDs, respectively.
			Then, the 3D receive wavevector ${\boldsymbol{\kappa}_{n, {\ell}^{\text{r}}}^{\text{r}} \in \mathbb{R}^{3 \times 1}}$ of the ${{\ell}^{\text{r}}}$-th receive channel path is expressed by
			\begin{equation}
				\boldsymbol{\kappa}_{n, {\ell}^{\text{r}}}^{\text{r}}= \frac{{2\pi }}{\lambda } \left[ {\cos {\theta ^{\text{r}}_{n,{{\ell}^{\rm{t}}}}}\cos {\phi ^{\text{r}}_{n,{{\ell}^{\rm{t}}}}},\cos {\theta^{\text{r}} _{n,{{\ell}^{\text{r}}}}} \sin {\phi ^{\text{r}}_{n,{{\ell}^{\text{r}}}}}, \sin {\theta ^{\text{r}}_{n,{{\ell}^{\text{r}}}}}} \right]^{\top},
			\end{equation}
			where ${ 1 \le {\ell}^{\text{r}} \le L^{\text{r}}}$.
			${\theta^{\text{r}} _{n,{{\ell}^{\text{r}}}}}$ and ${\phi^{\text{r}} _{n,{{\ell}^{\rm{t}}}}}$ are the elevation and azimuth AoAs, respectively.
            For the channel from the ${m}$-th MA to the ${n}$-th user, its transmit and receive field-response vectors (FPVs) based on the field-response channel model \cite{10304448} can be defined as 
			\begin{subequations}
				\begin{align}
				\boldsymbol{q}_{n}\left({\boldsymbol{t}_m}\right) &= 
				\left[ {{e^{j\rho _{n,1}^{\rm{t}}\left( {{\boldsymbol{t}_m}} \right)}}, \cdots ,{e^{j\rho _{n,{L^{\rm{t}}}}^{\rm{t}}\left( {{\boldsymbol{t}_m}} \right)}}} \right]^{\top} \in \mathbb{R}^{L^{\rm{t}} \times 1},\\
				\boldsymbol{f}_n\left({\boldsymbol{u}_{n}}\right) &=
				\left[ {{e^{j\rho _{n,1}^{\rm{r}}\left( {{\boldsymbol{u}_m}} \right)}}, \cdots ,{e^{j\rho _{n,{L^{\rm{r}}}}^{\rm{r}}\left( {{\boldsymbol{u}_m}} \right)}}} \right]^{\top} \in \mathbb{R}^{L^{\rm{r}} \times 1} .
				\end{align}
			\end{subequations}
			where ${\rho _{n,\ell^{\text{t}}}^{\rm{t}}\left( {{\boldsymbol{t}_m}} \right) = {\boldsymbol{t}_m^{\top}}\boldsymbol{\kappa}_{n, {\ell}^{\text{t}}}^{\text{t}}}$ and ${\rho _{n,\ell^{\text{r}}}^{\rm{r}}\left( {{\boldsymbol{u}_n}} \right) = {\boldsymbol{u}_n^{\top}}\boldsymbol{\kappa}_{n, {\ell}^{\text{r}}}^{\text{r}}}$ are the phase variations.

			The channel vector ${\mathbf{h}_{n}^{\text{ma}} \in \mathbb{C}^{M_{\text{ma}} \times 1}}$ between all receive and transmit channel paths from the BS to the ${n}$-th user can be expressed as 
			\begin{equation}
				\mathbf{h}_{n}^{\text{ma}} = \mathbf{Q}^{\top}_{n}\mathbf{\Sigma}_n\boldsymbol{f}_n\left(\boldsymbol{u}_n\right),
			\end{equation}
			where ${\mathbf{Q}_{n}=\left[{\boldsymbol{q}_n\left({\boldsymbol{t}_1}\right), \cdots, \boldsymbol{q}_n\left({\boldsymbol{t}_M}\right)}\right] \in \mathbb{C}^{L^{\rm{t}} \times M_{\text{ma}}}}$ is the transmit field-response matrix (FRM).
			${\mathbf{\Sigma}_n \in \mathbb{C}^{L^{\text{r}}\times L^{\text{t}}}}$ indicates the path-response matrix (PRM) and each element ${\left[ {{{\boldsymbol{\Sigma }}_n}} \right]_{\ell^{\text{t}}, \ell^{\text{r}}}}$ represents the channel response between the BS origin and the receiving origin, where the signal departures from the ${\ell^{\text{t}}}$-th transmit channel path and is received at the ${\ell^{\text{r}}}$-th receive channel path \cite{10416896}.
			Thus, the channel ${\mathbf{H}_{\text{ma}} \in \mathbb{C}^{M_{\text{ma}} \times N}}$ of the MA-enabled wireless system can be denoted as
			\begin{equation}
				\mathbf{H}_{\text{ma}} = \left[ {{\bf{h}}_1^{{\rm{ma}}}, \cdots ,{\bf{h}}_N^{{\rm{ma}}}} \right].
				\label{eq_channel4ma}
			\end{equation}
			
		\subsection{PA-enabled wireless system \label{sec_channel4pa}}
			In \ref{fig_system_model}(b), the BS is equipped with ${K_\text{wav}}$ dielectric waveguides and incorporates ${M_{\text{pa}}}$ PAs on each waveguide.
			All waveguides are assumed to be evenly aligned parallel to  the x-axis at the height of ${z_\text{wav}}$ \cite{11300296, 11267231}. 
			Meanwhile, each waveguide is assumed to have a length equal to that of the square area along the x-axis \cite{11223640}, i.e., ${L_{\text{wav}} = D_{\text{area}}}$.
			To ensure the fairness in the comparison, the number of flexible antennas in the two systems should satisfy
			${{K_{\text{wav}}}{M_{\text{pa}}} = {M_{\text{ma}}}}$, and the PA system should have the same number of candidate positions as the above MA system.
			Hence, ${I_{\text{pos}}}$ candidate positions are evenly distributed with antenna spacing ${
            L_{\text{wav}}/(I_{\text{pos}}-1)}$ \cite{11165763}.
			Let ${\mathcal{P}_{\text{pa}}^{\left({k}\right)} = \{ {{\bf{p}}_{k,1}^{{\rm{pa}}} \cdots ,{\bf{p}}_{k,{I_{{\rm{pos}}}}}^{{\rm{pa}}}} \}}$, ${\forall {\bf{p}}_i^{{\rm{pa}}} \in \mathbb{R}^{3\times 1}}$, denote the set of candidate positions on the ${k}$-th waveguide.
            For the ${m}$-th PA on the ${k}$-th waveguide, its 3D position can be defined as  ${\boldsymbol{t}_{k, m} = \left[ {{x_m^{\text{pa}}},{y_k^{\text{wav}}}, z_\text{wav}} \right]^{\top} \in \mathcal{P}_{\text{pa}}^{\left({k}\right)}}$, ${1 \leq k \leq K_{\text{wav}}}$, ${1 \leq m \leq M_{\text{pa}}}$.
			
			Then, the channel coefficient ${h_{k,m, n}}$ from the ${m}$-th PA on the ${k}$-th waveguide to the ${n}$-th user is represented as
			\begin{equation}
                h_{k,m, n} = \frac{{e^{ - j_{{\lambda _\text{g}}}^{2\pi }{{\left\| {{\boldsymbol{t}_{k,0}} - {\boldsymbol{t}_{k,m}}} \right\|}_2}}}}{{\sqrt {{M_{{\rm{pa}}}}} }} \frac{{\eta^{\frac{1}{2}} {e^{ - j\frac{{2\pi }}{\lambda }{{\left\| {{\boldsymbol{u}_n} - {\boldsymbol{t}_{k,m}}} \right\|}_2}}}}}{{{{\left\| {{\boldsymbol{u}_n} - {\boldsymbol{t}_{k,m}}} \right\|}_2}}},
            \end{equation}
			where $\eta^{\frac{1}{2}} = c/(4\pi f_c)$ with ${c}$ denoting the speed of light and ${f_c}$ denoting the carrier frequency.
            ${\lambda_g = \lambda/n_{\text{eff}}}$ is the carrier wavelength in the waveguide.
			${n_{{\text{eff}}}}$ denotes an effective refractive index of the waveguide.
			Thus, the channel ${\mathbf{H}_\text{pa} \in \mathbb{C}^{K_\text{wav} \times N}}$ of the PA-enabled wireless system can be given by
			\begin{equation}
				\mathbf{H}_\text{pa} = 
                \left[ {\begin{array}{*{20}{c}}
						{{h_{1,1}}}&{{h_{1,2}}}& \cdots &{{h_{1,N}}}\\
						{{h_{2,1}}}&{{h_{2,2}}}& \cdots &{{h_{2,N}}}\\
						\vdots & \vdots & \ddots & \vdots \\
						{{h_{K_\text{wav},1}}}&{{h_{K_\text{wav},2}}}& \cdots &{{h_{K_\text{wav},N}}}
				\end{array}} \right],
				\label{eq_channel4pa}
			\end{equation}
			where ${{h_{k,n}} = \sum\nolimits_{m = 1}^{{M_{{\text{pa}}}}} {{h_{k,m,n}}}}$ denotes the joint channel between ${M_{\text{pa}}}$ PAs on the ${k}$-th waveguide to the ${n}$-th user \cite{papanikolaou2025physical}.
	
	\subsection{Problem formulation}
        \subsubsection{Effective rank definition}
		To quantify and compare the spatial DoF between MA and PA-enabled wireless systems, we introduce the effective rank as a metric to evaluate the orthogonality of the channel ${\mathbf{H} \in \mathbb{C}^{M\times N}}$.
        ${{\mathop{\rm erank}\nolimits} \left( {\mathbf{H}} \right)}$ is defined based on the normalized singular value distribution of the channel matrix and is therefore invariant to any scalar scaling \cite{10217398}. 
        Since transmit power only introduces a multiplicative scaling factor to the channel matrix, it does not alter the normalized singular values. 
        This guarantees that ${{\mathop{\rm erank}\nolimits} \left( {\mathbf{H}} \right)}$ is independent of transmit power and depends only on the spatial characteristics of the channel \cite{10314137}.
		
        Hence, the effective rank between transmitting sources (i.e., MAs and PAs) and  receiving fields (i.e., users) can be expressed as \cite{roy2007effective}
		\begin{equation}
			{\mathop{\rm erank}\nolimits} \left( {\mathbf{H}} \right) = \exp ( { - \sum\nolimits_i {{{\tilde \sigma }_i}\ln {{\tilde \sigma }_i}} } ),
		\end{equation}
		where ${\tilde{\sigma_i} = \sigma_i/(\sum\nolimits_i \sigma_i)}$ is the ${i}$-th normalized singular value of ${\mathbf{H}}$.
        ${{\mathop{\rm erank}\nolimits} \left( {\mathbf{H}} \right)}$  ranges from ${0}$ to ${\min\left({M, N}\right)}$, increasing as the singular values grow more uniform \cite{11225901}.
 
		\subsubsection{MA system}
			Denote the selection vector as ${{\boldsymbol{\xi }}_{m} \in \left\{ {0,1} \right\}^{I_{\text{pos}} \times 1}}$ for the ${m}$-th MA.
			Its optimization objective is formulated to maximize ${{\mathop{\rm erank}\nolimits} \left( {{{\bf{H}}_{{\rm{ma}}}}} \right)}$ by selecting positions for ${M_\text{ma}}$ MAs on the 2D array plane, i.e.,  
			\begin{subequations}
				\label{eq_erank4ma}
				\begin{align}
					\mathop {\max }\limits_{{\boldsymbol{\xi }}_{m}}  \text{ }
					&{\mathop{\rm erank}\nolimits} \left( {{{\bf{H}}_{{\rm{ma}}}}} \right), \label{eq_erank4ma_obj}\\
					\text{s.t.} 
					\text{ }
					&{\left[ {{{\boldsymbol{\xi }}_{m}}} \right]_i} \in \left\{ {0,1} \right\}, \forall m, i, \label{eq_erank4ma_cons1} \\
					\text{ }
					&{\boldsymbol{\xi }}_m^{\top}{{\boldsymbol{\xi }}_{m'}} = 0, m  \ne m', \label{eq_erank4ma_cons2}
				\end{align}
			\end{subequations}
			where \eqref{eq_erank4ma_cons1} indicates that if ${\left[{{\boldsymbol{\xi }}_{m}}\right]_{i} = 1}$, the ${i}$-th candidate position is selected for the ${m}$-th MA (i.e., ${{\boldsymbol{t}_m} = \mathbf{p}_{i}^{\text{ma}}} \in \mathcal{P}_{\text{ma}}$); otherwise, it is not selected.

		\subsubsection{PA system}
			Denote the selection vector as ${{\boldsymbol{\xi }}_{k,m} \in \left\{ {0,1} \right\}^{I_{\text{pos}} \times 1}}$ for the ${m}$-th PA on the ${k}$-th waveguide.
			Its optimization objective is formulated to maximize ${{\mathop{\rm erank}\nolimits} \left( {{{\bf{H}}_{{\rm{pa}}}}} \right)}$ by selecting  positions for ${M_\text{pa}}$ PAs on ${K_\text{wav}}$ waveguides, i.e.,  
			\begin{subequations}
				\label{eq_erank4pa}
				\begin{align}
					\mathop {\max }\limits_{{\boldsymbol{\xi} _{k,m}}}   \text{ }
					&{\mathop{\rm erank}\nolimits} \left( {{{\bf{H}}_{{\rm{pa}}}}} \right), \label{eq_erank4pa_obj}\\
					\text{s.t.}
					\text{ }
					&{{{\left[ {{\boldsymbol{\xi }}_{k,m}} \right]}_i} \in \left\{ {0,1} \right\},\forall k,m,i}, \label{eq_erank4pa_cons1}\\
					\text{ }
					&\boldsymbol{\xi} _{k,m}^{\top}{\boldsymbol{\xi} _{k,m'}} = 0, m  \ne m',\label{eq_erank4pa_cons2}
				\end{align}
			\end{subequations}
			where \eqref{eq_erank4pa_cons1} indicates that if ${\left[{{\boldsymbol{\xi }}_{k, m}}\right]_{i} = 1}$, the ${i}$-th candidate position on the ${k}$-th waveguide is selected for the ${m}$-th PA (i.e., ${{\boldsymbol{t}_{k, m}} = \mathbf{p}_{k, i}^{\text{pa}}} \in \mathcal{P}^{\left({k}\right)}_{\text{pa}}$); otherwise, it is not selected.

            Note that the collision-free constraint \eqref{eq_erank4ma_cons2} ensures that multiple MAs do not occupy the same position on the 2D array plane.
            In contrast, the collision-free constraint \eqref{eq_erank4pa_cons2} guarantees that multiple PAs on the same waveguide do not occupy the identical position along that waveguide.
       		Then, both \eqref{eq_erank4ma} and \eqref{eq_erank4pa} are combinatorial optimization problems, which are generally NP-hard due to the binary nature of ${\boldsymbol{\xi}_m}$ and ${\boldsymbol{\xi}_{k, m}}$, and the non-convex collision-free constraints.
            These flexible antenna systems in practical implements are inherently dynamic rather than static, with input parameters such as user positions evolving over multi-time slots \cite{11020999}. 
            Since the propagation geometry depends on the positions of both flexible antennas and users, the channel ${\mathbf{H}_{\text{ma}}}$ or  ${\mathbf{H}_{\text{pa}}}$ also varies as user positions change at each time slot ${t}$, i.e., $\mathbf{H}_{\text{ma}}\left[{t}\right]$ or $\mathbf{H}_{\text{pa}}\left[{t}\right]$.
            Hence, these traditional one-shot optimization algorithms \cite{11226954, 11218873, 11164786, 11175694, 11184829, 11300296} designed for static scenarios need to be re-implemented over multi-time slots, resulting in prohibitively high computational complexity.

            Besides, since ${{\mathop{\rm erank}\nolimits} \left( {\mathbf{H}} \right)}$ is determined by the channels' spatial correlation between flexible antennas and users, which are inherently governed by their spatial distribution \cite{9930883}.
            Unlike traditional optimization algorithms that passively rely on instantaneous CSI, we model user spatial distribution via graph models, employ the graph neural network (GNN) for feature extraction, and integrate reinforcement learning to learn flexible antenna dynamic positioning for multi-time slot effective rank optimization. 

	\section{GRL-Based MA Dynamic Positioning Scheme \label{sec_gaiqn}}

		In this section, we focus on solving \eqref{eq_erank4ma} for dynamic positioning of multiple MAs over multi-time slots, where a novel GRL-based MA dynamic positioning scheme is proposed as shown in \reffig{fig_scheme4grl}.
		It covers the construction of the spatial distribution-based user graph model, the GRL framework, and the GAIQN algorithm.

	   \subsection{Spatial Distribution-Based User Graph Construction \label{sec_graph4grl}}
            To leverage the GNN for MA dynamic positioning design, a user graph model is built based on the spatial information of users' distribution in the square area. 
            At each time slot, it can be denoted by ${\mathcal{G}_{\text{ma}}=\left({\mathcal{N}, \mathcal{E}_{\text{ma}}}\right)}$, where $\mathcal{N} = \left\{1, 2, \cdots, N\right\}$ and ${\mathcal{E}}$ denote the set of vertices and undirected edges connecting these vertices, respectively.
    		An edge ${e_{n, n'}}$ between user ${n}$ and  ${n'}$ (${n \ne  n'}$) is established based on two aspects: 
            1) \textit{Euclidean Distance:} The Euclidean distance between user ${n}$ and ${n'}$ is expressed by ${d_{n, n'} = {\left\| {{{\boldsymbol{u}}_n} - {{\boldsymbol{u}}_{n'}}} \right\|_2}}$.
            Let  ${e_{n,n'}^{d}}$  denote an Euclidean distance-based edge.
            If ${d_{n, n'} \leq d_{\text{threshold}}}$, the edge is formed, i.e., ${e_{n,n'}^{d}=1}$; otherwise, ${e_{n,n'}^{d}=0}$. 
            2) \textit{Angular Direction:} The angular direction of the ${n}$-th user regarding the BS can be given by $\theta_n = \arctan((x_n - x_{\text{BS}})/(y_n - y_{\text{BS}}))$.
            Let  ${e_{n, n'}^{\theta}}$  denote the angular direction-based between user ${n}$ and user ${n'}$ .
            If ${\left| {{\theta _m} - {\theta _{m'}}} \right| \leq \theta_{\text{threshold}} }$, an edge is established, i.e.,  ${e_{n,n'}^{\theta}=1}$; otherwise, ${e_{n,n'}^{\theta}=0}$. 
    		Hence, for the adjacent matrix ${\mathbf{A}_t \in \left\{ {0,1} \right\}^{N \times N}}$ of the graph ${\mathcal{G}_{\text{ma}}}$ at time slot ${t}$, each element is represented by ${{\left[ {\mathbf{A}_t} \right]_{n, n'}} = {e_{n, n'}^{\theta}} \lor {e_{n, n'}^{d}}}$.

        \subsection{GRL Framework \label{sec_grl_framework}}
            \begin{figure}[!t]
                \centering
                \includegraphics[width=0.95\linewidth]{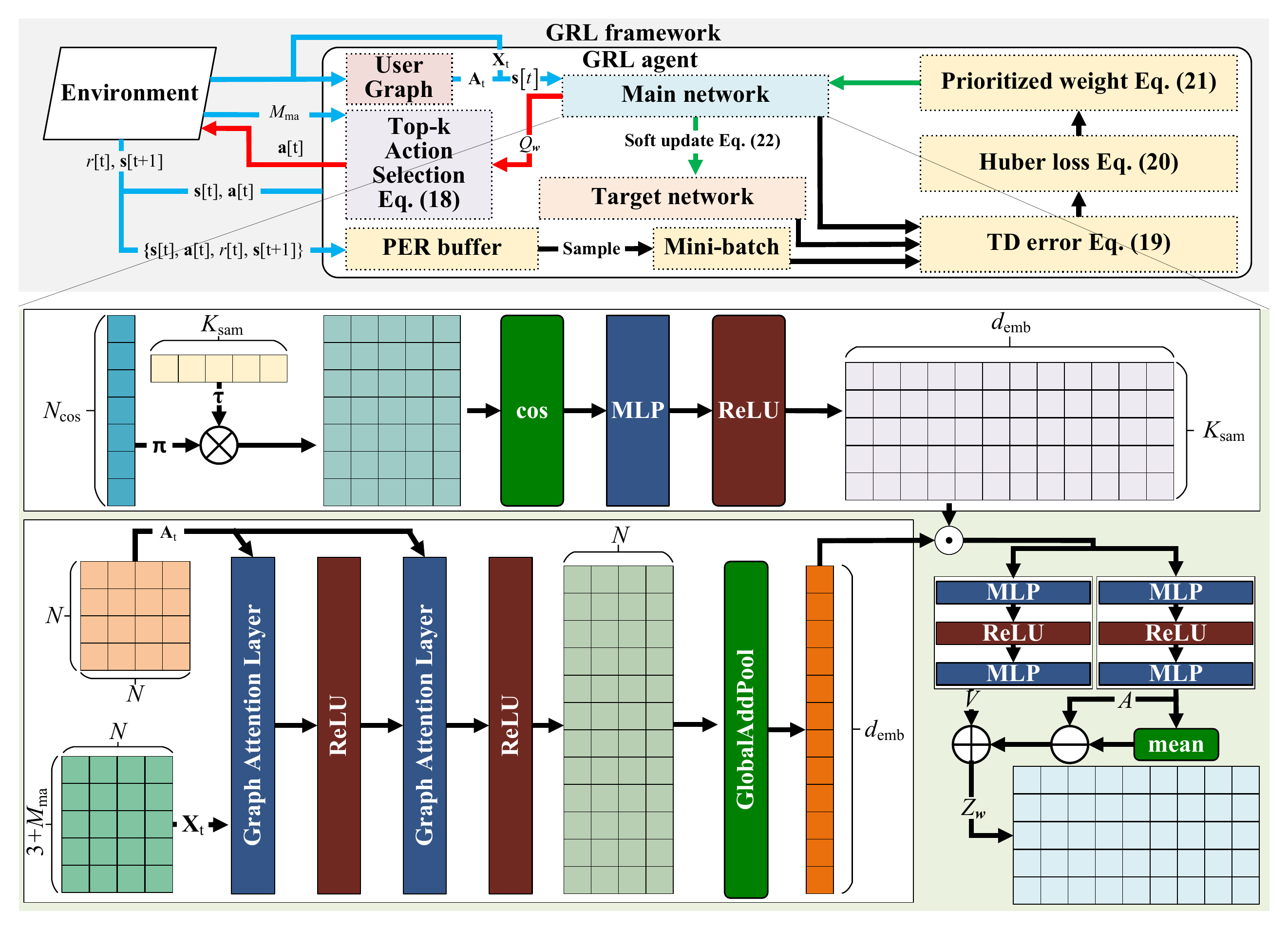}
                \caption{Flow chart of proposed GRL-based MA dynamic positioning scheme.}
                \label{fig_scheme4grl}
            \end{figure}
            
            %

        	\subsubsection{Agent Design}
        		The BS will act as a GRL agent, first collecting local state information from the unknown environment of the MA‑enabled wireless system.
        		At each time slot ${t}$, the BS then generates a joint action decision for all MAs and computes an immediate reward ${r_t}$ to evaluate the effectiveness of the decision.
        		Using the proposed GAIQN algorithm in \refsec{sec_algo4gaiqn}, the BS progressively learns a model‑free policy that maps observed states to optimize MA positions on the 2D array plane.

        	\subsubsection{Graph State}
        		Unlike the traditional RL framework, which only requires feature information as state, the current graph state ${\mathbf{s}\left[ t \right]}$ of the GRL agent will involve the features of users  (i.e., feature matrix ${\mathbf{X}_t}$) and the spatial relationship between users (i.e.,  adjacent matrix ${\mathbf{A}_t}$ based on \refsec{sec_graph4grl}) in the user graph  at time slot ${t}$.
        		For each user as vertex, its feature vector ${{\left[ {{{\mathbf{X}}_t}} \right]_{n, :}}}$ is defined by ${			{\left[ {{{\bf{X}}_t}} \right]_{n,:}} = \left[ {{{\bf{u}}_n}\left[ t \right]||{\bf{h}}_n^{{\rm{ma}}}\left[ t \right]} \right]}$,
        		where ${| {{{\left[ {{{\bf{X}}_t}} \right]}_{n, :}}} | = 3 + {M_{{\rm{ma}}}}}$. 
        		Accordingly, the current graph state ${\mathbf{s}\left[ t \right]}$ of the GRL agent can be expressed as ${{\mathbf{s}\left[ t \right]} = \left\{ {{{\mathbf{X}}_t},{{\mathbf{A}}_t}} \right\}}$.

        	\subsubsection{Action}
                Since the decision variables are MAs' positions in \eqref{eq_erank4ma}, the GRL agent will execute a joint action ${\mathbf{a}\left[ t \right]  \in \mathbb{Z}^{M_{\text{ma}}}}$ (i.e., it is a  multidimensional discrete action) to select positions for all MAs.
        		The ${m}$-th MA corresponding action  is expressed by ${{a_m}\left[ t \right] \in {\mathcal{A}} \triangleq \left\{ {1,2, \cdots ,{I_{{\rm{pos}}}}} \right\}}$, where ${\mathcal{A}}$ denotes a discrete action space for each MA with size ${I_{{\rm{pos}}}}$. 
        		Furthermore, the relationship between action ${ {a_m}\left[ t \right]}$ and decision variable ${\boldsymbol{\xi}_{m}}$ is as ${[ \boldsymbol{\xi}_{m} ]_{i \coloneqq {a_m}\left[ t \right]}}$ satisfying \eqref{eq_erank4ma_cons1}.
        		Thus, the joint action ${\mathbf{a}\left[ t \right]}$ at time slot ${t}$ can be expressed by ${			\mathbf{a}\left[ t \right]= \left[ {{a_1}\left[ t \right],{a_2}\left[ t \right], \cdots ,{a_{{M_{{\rm{ma}}}}}}\left[ t \right]} \right]}$.

        	\subsubsection{Reward Function}
        		After executing ${\mathbf{a}\left[ t \right]}$ at time slot ${t}$, the reward function ${\mathcal{R}_{\text{ma}}}$ will evaluate the quality of selected MA positions in terms of the optimization objective \eqref{eq_erank4ma_obj}, i.e., effective rank ${{\mathbf{H}}_{\text{ma}}\left[{t}\right]}$ based on \eqref{eq_channel4ma}. 
        		Hence, the immediate reward ${r\left[{t}\right]}$ in the built GRL framework is defined as
        		\begin{equation}
       				{r\left[{t}\right]} = \mathcal{R}_{\text{ma}}\left( {{{\mathbf{s}}\left[ t \right]},{{\mathbf{a}}\left[ t \right]}} \right) = {\mathop{\rm erank}\nolimits} \left( {{{\mathbf{H}}_{\text{ma}}\left[{t}\right]}} \right) + {\mathcal{C}}_{\text{ma}}\left[ t \right],\\
        			\label{eq_reward4ma}
        		\end{equation}
        		where ${{{\mathcal{C}}_{{\rm{ma}}}}\left[ t \right] =  - \alpha \sum\nolimits_{m = 1}^{{M_{{\rm{ma}}}}} {\sum\nolimits_{m' = m + 1}^{{M_{{\rm{ma}}}}} {{\bf{\xi }}_m^{\top}{{\bf{\xi }}_{m'}}} } \leq 0}$ is a penalty to ensure \eqref{eq_erank4ma_cons2}, based on the number of collisions between multiple MAs on the 2D array plane.
                ${\alpha}$ is a hyperparameter used to regulate the intensity of penalty.
		
        \subsection{GAIQN Algorithm \label{sec_algo4gaiqn}}
            To enable the GRL agent to gradually obtain better optimal positions and ensure collision-free between multiple MAs, we propose a GAIQN-based top-k MA positioning algorithm (labeled as the GAIQN algorithm) consisting of the following three components:
    		\subsubsection{Graph Attention Implicit Quantile Network Structure}
    			As shown in \reffig{fig_scheme4grl}, the graph attention \cite{velivckovic2017graph} is introduced to effectively handle the input graph state to achieve the graph embedding: ${\mathbb{R}^{3+M_{\text{ma}}} \rightarrow \mathbb{R}^{d_{\text{emb}}}}$.
    			A scoring function ${f_{\text{e}}:\mathbb{R}^{d_{\text{in}}}\times \mathbb{R}^{d_{\text{out}}} \rightarrow \mathbb{R}}$ computes a score for every \textit{edge} ${e_{n, n'}}$, which indicates the importance of the features of the neighbor ${n'}$ to the vertex ${n}$:
    			\begin{equation}
                    f_{\text{e}}\left({\mathbf{x}^{\text{in}}_n, \mathbf{x}^{\text{in}}_{n'}}\right) = {\mathop{\rm LeakyReLU}\nolimits} ({\boldsymbol{\beta}^{\top} \odot \left[{\mathbf{W}\mathbf{x}^{\text{in}}_{n} || \mathbf{W}\mathbf{x}^{\text{in}}_{n'}}\right] }),
    			\end{equation}
    			where ${\boldsymbol{\beta} \in \mathbb{R}^{2d_{\text{out}}}}$, ${\mathbf{W} \in \mathbb{R}^{d_{\text{out}}\times d_{\text{in}}}}$ are learned.
    			${d_\text{in}}$ and ${d_\text{out}}$ are the dimensions of the input and output, respectively.
                ${{\mathop{\rm LeakyReLU}\nolimits} \left(  \cdot  \right)}$ is a nonlinear activation function.
    			These attention scores will be normalized across all neighbors ${n' \in {\mathcal{N}_n} \triangleq \{ {\tilde n \in \mathcal{N} | {{{\left[ \mathbf{A} \right]}_{n,\tilde n}} = 1} } \}}$ using the softmax function. 
                It can be defined as
    			\begin{equation}
                    \beta_{n, n'} 
                    = \frac{\exp\left({{f_{\text{e}}\left({\mathbf{x}^{\text{in}}_n, \mathbf{x}^{\text{in}}_{n'}}\right)}}\right)}{ {\textstyle \sum_{n''\in \mathcal{N}_{n}} \exp\left({f_{\text{e}}\left({\mathbf{x}^{\text{in}}_n, \mathbf{x}^{\text{in}}_{n''}}\right)}\right)} }.
    			\end{equation}
            
                Consequently, the graph attention computes a weighted average of the transformed features of the neighbor vertices as the new representation of the ${n}$-th vertex using the normalized attention coefficients, which can be defined as 
                \begin{equation}
                    \mathbf{x}^{\text{out}}_{n} = \rm{ReLU}\left({\textstyle \sum_{n'\in \mathcal{N}_{n}} \beta_{n, n'}  \odot \mathbf{W}\mathbf{x}^{\text{in}}_{n'}}\right),
                    \label{eq_gat}
                \end{equation}
                where ${{\mathop{\rm ReLU}\nolimits} \left(  \cdot  \right)}$ is a nonlinear activation function.
                Meanwhile, the graph-level embedded vector ${\mathbf{z}^{\text{g}} \in \mathbb{R}^{d_{\text{emb}}}}$ is computed by adding node-level embedded vectors, which is defined by ${{{\bf{z}}^{\rm{g}}} = \sum\nolimits_{n = 1}^N {{\bf{x}}_n^{{\rm{out}}}} }$.
                Since this structure needs to select positions for MAs, the implicit quantile approach \cite{dabney2018implicit} is adopted to estimate the distribution of all candidate positions on the 2D array plane.
                Instead of the expected return ${Q\left({s, a}\right)}$, the return distribution ${Z\left({s, a}\right)}$ is expressed using a quantile function ${Z_{\tau} \coloneqq F^{-1}_{Z}\left({\tau}\right)}$,
                where ${F^{-1}_{Z}\left({\tau}\right)}$ is the quantile function at ${\tau \sim U\left({\left[{0, 1}\right]}\right)}$ for the random variable ${Z}$.
                In \reffig{fig_scheme4grl}, the \textit{cosine} embedding ${{\boldsymbol{\Phi }}\left( {\boldsymbol{\tau }} \right) \in \mathbb{R}^{K_{\text{sam}} \times d_{\text{emb}}}}$ is represented as
                \begin{equation}
                    {\left[ {{\boldsymbol{\Phi }}\left( {\boldsymbol{\tau }} \right)} \right]_{k,:}} = {\rm{MLP}}\left( {\left[ {\cos \left( {1\pi {\tau _k}} \right), \cdots ,\cos \left( {{N_{\cos }}\pi {\tau _k}} \right)} \right]} \right),
                \end{equation}
                where ${{\boldsymbol{\tau }} = {\left[ {{\tau _1},{\tau _2}, \cdots ,{\tau _{{K_{{\rm{sam}}}}}}} \right]^{\top}}}$. 
                The obtained quantile information will further be modulated into the graph embedding vector ${\mathbf{z}^{\rm{g}}}$ to derive different value distribution representations from a graph state, thereby implicitly modeling the entire return distribution.
                Hence, it can be calculated by
                \begin{equation}
                    {\mathbf{Z}}_{\boldsymbol{\tau}} ^{\rm{g}} = {{\mathbf{z}}^{\rm{g}}} \odot \boldsymbol{\Phi} \left( \boldsymbol{\tau}  \right) \in \mathbb{R}^{K_{\text{sam}} \times d_{\text{emb}}}.
                \end{equation}			
                In addition, the dueling architecture is adopted to boost learning efficiency and stability, 
                which can be expressed by
                \begin{equation}
                    Z_{\boldsymbol{w}}\left( {\mathbf{s}, \mathbf{a}, \boldsymbol{\tau} } \right) =V\left( {\mathbf{s},\boldsymbol{\tau} } \right) + ( {A\left( {\mathbf{s}, \mathbf{a}, \boldsymbol{\tau} } \right) - \frac{{\sum\nolimits_{a'} {A\left( {{\bf{s}},a',{\boldsymbol{\tau }}} \right)} }}{{\left| {{I_{pos}}} \right|}}} ),
                \end{equation}
                where ${\boldsymbol{w}}$ denotes the network parameters that need to be trained. ${V}$ and $A$ are value and advantage streams, respectively.
                This decomposition of the dueling architecture allows the agent to learn the value of a state independently of the specific action taken.
                This approximation is obtained by drawing ${K_{\text{sam}}}$ levels ${{\tau _1},{\tau _2}, \cdots ,{\tau _{{K_{{\rm{sam}}}}}}}$ uniformly and independently from the ${\left({0, 1}\right)}$ interval and averaging the output of the network at these levels: ${{Q_{\boldsymbol{w}}}\left( {\mathbf{s}, \mathbf{a}} \right) \approx \mathbb{E}_{\tau_{k} \sim U\left({\left[{0, 1}\right]}\right)} \left[{Z_{\boldsymbol{w}}\left( {\mathbf{s}, \mathbf{a}, \boldsymbol{\tau} } \right)}\right]}$.

                \begin{figure*}[!t]
                    \begin{equation}
                        \delta _{i,t}^{\tau_{k} ,\tau_{k'}} = r^{\left( i \right)}\left[ t \right] + \gamma { {{Z_{\boldsymbol{w}'}^{\left({\tau_{k'}}\right)}}( {{\bf{s}}^{\left( i \right)}\left[ {t + 1} \right],\mathop {\rm{argtopk} }\limits_{{\bf{a}}^{\left( i \right)}\left[ {t + 1} \right]} {Z_{\boldsymbol{w}}^{\left({\tau_{k'}}\right)}}( {{\bf{s}}^{\left( i \right)}\left[ {t + 1} \right],{\bf{a}}^{\left( i \right)}\left[ {t + 1} \right],\boldsymbol{\tau} } ),\boldsymbol{\tau} } )} } - { {{Z_{\boldsymbol{w}}^{\left({\tau_{k'}}\right)}}( {{\bf{s}}^{\left( i \right)}\left[ {t} \right],{\bf{a}}^{\left( i \right)}\left[ {t} \right],\boldsymbol{\tau} } )} }
                        \tag{19}
                        \label{eq_td4ma}
                    \end{equation}
                    \rule{\textwidth}{0.4pt} 
                \end{figure*}
                
            \subsubsection{Top-k Action Selection for MA system\label{sec_topk}}
        		To effectively avoid multiple MAs from simultaneously occupying the same position, i.e., satisfying \eqref{eq_erank4ma_cons2}, a top-k action selection method is designed for the MA system. 
        		Unlike traditional action masking mechanisms, which rely on prior knowledge of interaction conflicts, the proposed top‑k action selection method inherently ensures multiple non‑repetitive Q‑values. 
        		Meanwhile, to balance the exploration and exploitation, the ${\varepsilon}$-greedy factor ${\varepsilon \left( t \right) \in \left( {0,1} \right)}$ is also introduced into this method.
        		At each time slot ${t}$, the GRL agent randomly generates ${M_{\text{ma}}}$ non-repeating actions with probability ${\varepsilon \left( t \right)}$; otherwise, with probability ${1 - \varepsilon \left( t \right)}$, it exploits the joint action ${\mathbf{a}\left[{t}\right]}$, i.e.,
        		\begin{equation}
                    \mathbf{a}\left[{t}\right] = 
                    \begin{cases}
                        \arg {\rm{top}}{{\rm{k}}_{\mathbf{a}}}\left( {{Q_{\boldsymbol{w}}}\left( {\mathbf{s}\left[ t \right], \mathbf{a}} \right), M_{\text{ma}}} \right), &{\text{if }} 1 - \varepsilon \left( t \right),\\ 
                        {\text{random},}&{\text{otherwise.}} 
                    \end{cases}
                    \label{eq_action_selection4ma}
        		\end{equation}

            \subsubsection{Parameter Update}
                Analogous to the conventional Bellman optimality equation of the Markov decision process, the value distributional Bellman optimality equation can be given as  ${				\mathcal{T}^{*}Z\left({\mathbf{s}, \mathbf{a}}\right) \overset{D}{=} \mathcal{R}\left({\mathbf{s}, \mathbf{a}}\right) + \gamma {Z }\left( {\mathbf{s}',\mathop {\arg \max }_{\mathbf{a}'} \mathbb{E}\left[ {{Z}\left( {\mathbf{s}',\mathbf{a}'} \right)} \right]} \right).}$ \cite{10540320}.
                Thus, its sampled temporal difference (TD) error between the ${k}$-th and ${k'}$-th samples of the ${i}$-th transition ${\left({{{\bf{s}}^{\left( i \right)}}\left[ t \right],{{\bf{a}}^{\left( i \right)}}\left[ t \right],{r^{\left( i \right)}}\left[ t \right],{{\bf{s}}^{\left( i \right)}}\left[ {t + 1} \right]}\right)}$ at time slot ${t}$ can be expressed in \eqref{eq_td4ma}.
                Then, these quantile estimates are trained using the \textit{Huber} loss, with threshold ${\kappa}$,
                \setcounter{equation}{19}
                \begin{equation}
                \mathcal{H}_{\kappa}\left({u}\right) = 
                \begin{cases}
                	\frac{1}{2} u^2, &\text{if } \left|{u}\right| \leq \kappa,\\
                	\kappa\left({\left|{u}\right| - \frac{1}{2}\kappa}\right), &\text{otherwise}.
                \end{cases}
                \end{equation}
			
                Hence, the loss function of the ${i}$-th transition at time slot ${t}$ can be defined as  ${\mathscr{L}^{(i)}(\boldsymbol{w}) = \sum\nolimits_{k=1}^{K_{\text{sam}}} \sum\nolimits_{k'=1}^{K'_{\text{sam}}}\rho_{\tau_k}^{\kappa}(\delta_{i, t}^{\tau_{k}, \tau_{k'}})/K'_{\text{sam}}}$,
                where  ${\rho_{\tau}^{\kappa}(u) = |\tau - \mathbb{I}_{\{u < 0\}}|\mathcal{H}_{\kappa}(u)/\kappa}$.
    			Besides, the prioritized experience replay (PER) buffer ${\mathbb{D}}$ is utilized, which employs a stochastic sampling method to select transitions based on their relative priorities.
    			The loss function of prioritized weight can be expressed by
                \begin{equation}
                    \hat{\mathscr{L}}^{\left( i \right)}\left({\boldsymbol{w}}\right) = \left({\frac{1}{\left|{\mathbb{D}}\right|}  \cdot \frac{1}{P\left({i}\right)}}\right)^{\beta} \mathscr{L}^{\left( i \right)}\left({\boldsymbol{w}}\right),
                    \label{eq_per_loss4gaiqn}
                \end{equation} 
                where $P(i) = (|\delta_{i,t} |+ \epsilon)/(\sum\nolimits_{i'}(|\delta_{i', t}|+ \epsilon)$  is the probability of sampling transition ${i}$. ${\delta _{i,t}}$ is the TD error of the ${i}$-th transition. ${\epsilon}$ denotes a small positive constant that avoids zero-error transitions being ignored.
                Then, the target network above uses a slow-moving update rate, parameterized by ${\tau_{\text{soft}} \in \left({0, 1}\right)}$, which can be expressed by
                \begin{equation}
                    \boldsymbol{w}' \leftarrow {\tau _{{\rm{soft}}}}\boldsymbol{w} + \left( {1 - {\tau _{{\rm{soft}}}}} \right)\boldsymbol{w}'.
                    \label{eq_soft_update4ma}
                \end{equation}
			
    			In summary, the workflow of the GAIQN algorithm is summarized in \refalg{algo_GAIQN}.

                \begin{algorithm}[!t]
                \caption{GAIQN Algorithm.}
                \label{algo_GAIQN}
                    \begin{algorithmic}[1]
                        \STATE Start the simulation system for the MA-enabled wireless system with  ${M_{\text{ma}}}$ MAs and ${N}$ users;
                        \STATE Initialize the parameters ${\boldsymbol{w}}$ randomly;
                        \STATE Initialize the parameters ${\boldsymbol{w}'}$: ${\boldsymbol{w}'  \leftarrow \boldsymbol{w}}$;
                        \STATE \textbf{for} \textit{episode\_idx}$ = \left\{ {1,2, \cdots ,N_{\text{episode}}} \right\}$ \textbf{do}
                        \STATE \hspace{0.3cm} Reset positions of MAs and users;
                        \STATE \hspace{0.3cm} \textbf{for} time slot $ t = \left\{ {1,2, \cdots , T} \right\}$ \textbf{do}
                        \STATE \hspace{0.3cm} \hspace{0.3cm} 
                        \begin{minipage}[t]{.9\linewidth}
                        The agent obtains the current state ${\mathbf{s}\left[{t}\right]}$ and then generates an action ${\mathbf{a}\left[{t}\right]}$ based on \eqref{eq_action_selection4ma};
                        \end{minipage}
                        \STATE \hspace{0.3cm} \hspace{0.3cm} Observe an immediate reward ${r_t}$ based on \eqref{eq_reward4ma};
                        \STATE \hspace{0.3cm} \hspace{0.3cm} Get the next state ${\mathbf{s}\left[{t+1}\right]}$ from the environment;
                        \STATE \hspace{0.3cm} \hspace{0.3cm} Store the transition  ${\left({{\mathbf{s}\left[{t}\right]},{\mathbf{a}\left[{t}\right]},{r\left[{t}\right]},{\mathbf{s}\left[{t+1}\right]}} \right)}$ into ${\mathbb{D}}$;

                        \STATE \hspace{0.3cm} \textbf{end} \textbf{for}				
                        \STATE \hspace{0.3cm}
                        \begin{minipage}[t]{.9\linewidth}
                        Sample a mini-batch of size ${I_{\text{batch}}}$ from ${\mathbb{D}}$; 
                        \end{minipage}
                        \STATE \hspace{0.3cm}
                        \begin{minipage}[t]{.95\linewidth}
                        For each transition ${i \in \left({1, 2, \cdots, I_{\text{batch}}}\right)}$, calculate ${\hat{\mathscr{L}}^{\left( i \right)}\left({\boldsymbol{w}}\right)\textbf{}}$ based on \eqref{eq_per_loss4gaiqn} to update ${\boldsymbol{w}}$.
                        \end{minipage}
                        \STATE \hspace{0.3cm}
                        \begin{minipage}[t]{.8\linewidth}
                        Soft update ${\boldsymbol{w}'}$ based on \eqref{eq_soft_update4ma}.
                        \end{minipage}
                        \STATE \textbf{end} \textbf{for}
                        \STATE \textbf{return:} the parameters ${\boldsymbol{w}}$ of main network.
                    \end{algorithmic}
                \end{algorithm}

            \subsection{Complexity Analysis}
                \subsubsection{Training Stage}
                    The computational complexity of the proposed GAIQN algorithm mainly involves four parts.
                    \begin{itemize}
                        \item         \textbf{Network Initialize:} 
                        This phase involves the initialization of network parameters for graph attention, implicit quantile, and dueling architecture components of GAIQN. Specifically, the computational complexity is expressed by
                        $ \mathcal{O}\left({\left|{\boldsymbol{w}}\right| + \left|{\boldsymbol{w}'}\right| }\right) $, 
                        where ${\mathcal{O}\left({\left|{\boldsymbol{w}'}\right|}\right) = \mathcal{O}\left({\left|{\boldsymbol{w}}\right|}\right)}$ and is specifically represented by  ${            \mathcal{O}\left({\sum\nolimits_{l = 1}^{{L_{\text{emb}}}} {d_{{\rm{in}}}^{\left( l \right)}} d_{{\rm{out}}}^{\left( l \right)} + N_{\cos}d_{\text{emb}}+ 2\sum\nolimits_{l = 1}^{{L_{{\rm{hid}}}}} {F_{{\rm{in}}}^{\left( l \right)}} F_{{\rm{out}}}^{\left( l \right)}}\right)}$.
                        
                        \item         \textbf{Action Sampling:}
                        Actions are generated based on the current graph state, with the computational complexity given by
                        ${\mathcal{O}\left({N_{\text{episode}} T \left|{\hat{\boldsymbol{w}}}\right|}\right)}$, where ${\mathcal{O}\left({\left|{\hat{\boldsymbol{w}}}\right|}\right) }$ denotes the computational complexity of network inference.
                        It can be specifically represented by ${\mathcal{O}(\sum\nolimits_{l = 1}^{{L_{{\rm{emb}}}}} {( {Nd_{{\rm{in}}}^{\left( l \right)} + \left| {{\mathcal{E}_{{\rm{ma}}}}} \right|} )} d_{{\rm{out}}}^{\left( l \right)} + N{d_{{\rm{emb}}}} +{K_{{\rm{sam}}}} (2{N_{\cos }}}$ 
                        ${ +{N_{\cos }}{d_{{\rm{emb}}}} + {d_{{\rm{emb}}}}) +2\sum\nolimits_{l = 1}^{{L_{{\rm{hid}}}}} {F_{{\rm{in}}}^{\left( l \right)}F_{{\rm{out}}}^{\left( l \right)}}+ {F_{{\rm{hid}}}}(I_{\text{pos}}+1) }$ 
                        ${+ {I_{{\rm{pos}}}})}$, 
                        where ${L_{\text{emb}}}$ and ${L_{\text{hid}}}$ denote the number of graph attention layers in graph embedding processing and MLP in dueling architecture, respectively.
                        
                        \item        \textbf{PER Buffer Collection}: The complexity of collecting transitions in the PER buffer is ${\mathcal{O}\left({N_{\text{episode}}\left({TV+ I_{\text{batch}}}\right)}\right)}$, where $\mathcal{O}\left(I_{\text{batch}}\right)$ and ${\mathcal{O}\left(V\right)}$ are the computational complexity of calculating priority of transitions and each interaction with the environment, respectively. 
                        \item        \textbf{Network Update}: Its computational complexity can be calculated as
                        ${\mathcal{O}\left({2N_{\text{episode}}\left| \boldsymbol{w} \right|}\right)}$.
                    \end{itemize}

                   In the training stage, the space complexity of the GAIQN is  ${\mathcal{O}\left({2\left|{\boldsymbol{w}}\right| + \left|{\mathbb{D}}\right|\left({2N\left({3+M_{\text{ma}}+N}\right) + M_{\text{ma}} + 1}\right)}\right)}$, where ${\left|{\mathbb{D}}\right|}$ represents the size of the PER buffer.
                  The space complexity is determined by the storage required for the neural network parameters and the data structures of the replay buffer, which stores the transitions. 

                \subsubsection{Execution Stage}     
                    The computational complexity of the GAIQN is ${\mathcal{O}\left({\left|{\hat{\boldsymbol{w}}}\right|}\right)}$, which is primarily due to the main network inferring actions based on the current graph state.
                    Besides, the space complexity during the execution stage is ${\mathcal{O}\left({\left|{\boldsymbol{w}}\right|}\right)}$, as the main network parameters must reside in memory to facilitate action selection for the GRL agent.

        \section{MAGRL-Based PA Dynamic Positioning Scheme \label{sec_magaqn}}
            In this section, we focus on solving \eqref{eq_erank4pa} for dynamic positioning of multiple PAs over multi-time slots, where a novel MAGRL-based PA dynamic positioning scheme is proposed as shown in \reffig{fig_scheme4magrl}.
            It covers the construction of  waveguide region-based user graph model, the MAGRL framework, and the MAGAQN algorithm.
            
        	\subsection{Waveguide Region-Based User Graph Construction}
        		Unlike all MAs implemented on the single BS, the PA elements are distributed on different waveguides according to \refsec{sec_system_model}.
        		A graph construction method based on K-means clustering is designed effectively to utilize the spatial information between waveguides and users, where the clusters' number is equal to the number of PAs on the ${k}$-th waveguide.
        		Here, the region formed between the ${\left(k-1\right)}$-th and ${\left(k+1\right)}$-th waveguides is the ${k}$-th waveguide region. 
        		Let ${\mathcal{N}_{k}^{\text{wav}} = \left\{ {n \in N\left| {y_{k - 1}^{{\rm{wav}}} < {y_n} \le y_{k + 1}^{{\rm{wav}}}} \right.} \right\}}$ denote the set of users within the ${k}$-th waveguide region.
        		After clustering, the 2D position of the ${\tilde{m}}$-th cluster within the ${k}$-th waveguide region is defined as ${{{\bf{c}}_{k,\tilde{m}}} = \left[ {{x_{k,\tilde{m}}},{y_{k,\tilde{m}}}} \right]}$.
        		${\mathcal{N}_{k}^{\text{wav}}}$ will be divided into multiple subsets ${\mathcal{N}_{k, \tilde{m}}^{\text{cluster}}}$, i.e., ${\mathcal{N}_{k, 1}^{\text{cluster}} \cup \cdots \cup \mathcal{N}_{k, M_{\text{ma}}}^{\text{cluster}} = \mathcal{N}_{k}^{\text{wav}}}$.
                Then, the user graph ${\mathcal{G}_{\text{pa}}^{k}}$ within the ${k}$-th waveguide region can be defined as
                ${\mathcal{G}_{\text{pa}}^{k} =\left({\mathcal{N}_{k}^{\text{wav}}, \mathcal{E}_{\text{pa}}^{k}}\right)}$
                where ${\mathcal{N}_{k}^{\text{wav}}}$ and ${\mathcal{E}_{\text{pa}}^{k}}$  are the set of vertices and undirected edges, respectively.
        		At each time slot, an edge ${e_{k, n, n'}}$ between users within the ${k}$-th waveguide region is established based on two aspects:
                1) \textit{Euclidean Distance:} The Euclidean distance between user ${n}$ and ${n'}$ in the ${\tilde{m}}$-th cluster is represented as  ${d_{\tilde{m}, n, n'}^{d} = {\left\| {{{\boldsymbol{u}}_n} - {{\boldsymbol{u}}_{n'}}} \right\|_2}}$, ${\forall n, n' \in \mathcal{N}_{k, \tilde{m}}^{\text{cluster}}}$.
                Let ${e_{k, n,n'}^{d}}$ denote the Euclidean distance-based edge within the ${k}$ waveguide region.
                If ${d_{\tilde{m}, n, n'}^{d}} \leq d_{\text{threshold}}$, the edge is formed, i.e., ${e_{k, n,n'}^{d}=1}$; otherwise, ${e_{k, n,n'}^{d}=0}$. 
                2) \textit{Angular Direction:} The angular direction of the ${n}$-th user in the ${\tilde{m}}$-th cluster of the ${k}$ waveguide region is defined as ${\theta_{\tilde{m}, n} = \arctan((x_n - x_{k, \tilde{m}})/(y_n - y_{k,\tilde{m}}))}$.
                Let  ${{e_{k, n,n'}^{\theta}}}$ denote an angular direction-based edge  in the ${k}$ waveguide region.
                If ${\left| {{\theta _{\tilde{m}, n}} - {\theta _{\tilde{m}, n'}}} \right| \leq \theta_{\text{threshold}} }$, an edge is established, i.e.,  ${{e_{k, n,n'}^{\theta}}=1}$; otherwise, ${{e_{k, n,n'}^{\theta}}=0}$. 
        		Thus, for the adjacent matrix ${\mathbf{A}^{k}_t \in \left\{ {0,1} \right\}^{ \left| {\mathcal{N}_{k}^{\text{wav}}} \right| \times \left| {\mathcal{N}_{k}^{\text{wav}}} \right|}}$ of the graph
                 ${\mathcal{G}_{\text{pa}}^{k}}$ in the ${k}$-th waveguide region at time slot ${t}$, each element is expressed as ${{\left[ {\mathbf{A}^{k}_t} \right]_{n, n'}} = {e_{n, n'}^{\theta}} \lor {e_{n, n'}^{d}}, \forall n,n' \in  \mathcal{N}_{k}^{\text{wav}}}$.

            \subsection{MAGRL Framework}

                \begin{figure}[!t]
                    \centering
                    \includegraphics[width=0.95\linewidth]{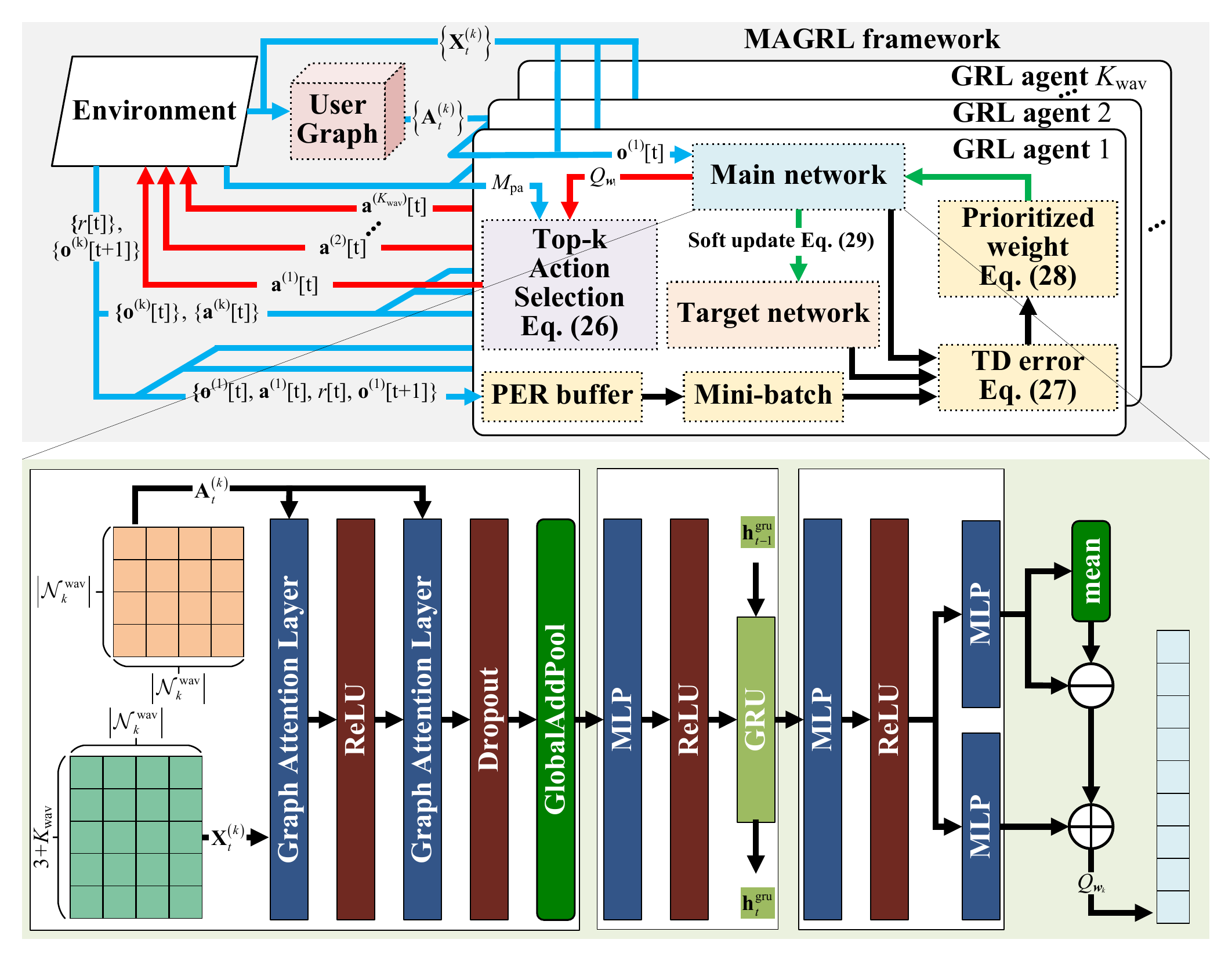}
                    \caption{Flow chart of proposed MAGRL-based PA dynamic positioning scheme.}
                    \label{fig_scheme4magrl}
                \end{figure}
   
        		\subsubsection{Agent Design}
        			Since multiple waveguides attempt to select positions for their PAs, each waveguide will act as a GRL agent and interact with the unknown environment to gain experiences, which are then used to direct its policy design.
        			Multiple agents collectively explore the environment and refine candidate positions based solely on their own observations and received information \cite{11320569}.
        			Hence, it will be turned into a fully cooperative one through using the same reward for all GRL agents, in the interest of global effective rank \eqref{eq_erank4pa_obj}.
			
        		\subsubsection{State and Graph Observation}
        			At each time slot ${t}$, given environment state ${S_t}$, each GRL agent ${k}$ will receive a graph observation ${\mathbf{o}^{\left({k}\right)}\left[{t}\right]}$ of the environment, determined by the observation function ${\mathscr{O}}$ as ${\mathbf{o}^{\left({k}\right)}\left[{t}\right] = \mathscr{O}\left({S_t, k}\right)}$, and then takes an action ${\mathbf{a}^{\left({k}\right)}\left[{t}\right]}$.
        			The graph observation ${\mathbf{o}^{\left({k}\right)}\left[{t}\right]}$ in the ${k}$-th waveguide region includes a feature matrix ${\mathbf{X}^{\left({k}\right)}_t}$ and an adjacent matrix ${\mathbf{A}^{\left({k}\right)}_t}$.
        			For each user as a vertex in the ${k}$-th waveguide region, its feature vector  is defined as ${				[{\mathbf{X}^{\left({k}\right)}_t}]_{n, :} =  {{{\bf{u}}_n}\left[ t \right]||[{\mathbf{H}^{\top}_{\text{pa}}[{t}]}]_{n, :}}, \forall n \in  \mathcal{N}_{k}^{\text{wav}}}$,
        			where ${|{[{\mathbf{X}^{\left({k}\right)}_t}]_{n, :}}| = 3 + K_{\text{wav}}}$.
        			Then, the current graph observation ${\mathbf{o}^{\left({k}\right)}\left[{t}\right]}$  of the ${k}$-th GRL agent at time slot ${t}$ is represented by ${{\mathbf{o}^{\left({k}\right)}\left[ t \right]} = \{ {{{\mathbf{X}}^{\left({k}\right)}_t},{{\mathbf{A}}^{\left({k}\right)}_t}} \}}$.

        		\subsubsection{Action} 
        			At time slot ${t}$, each GRL agent ${k}$ will execute a joint action ${\mathbf{a}^{\left({k}\right)}\left[{t}\right] \in \mathbb{Z}^{M_{\text{pa}}}}$ to select candidate positions for its PAs, which is a multidimensional discrete action.
        			According to \refsec{sec_channel4pa}, the ${m}$-th PA on the ${k}$-th waveguide corresponding action can be defined by $a_{k,m}\left[{t}\right] \in \mathcal{A} \triangleq \left\{{1, 2, \cdots, I_{\text{pos}}}\right\}$.
        			Meanwhile, the relationship between action ${a_{k,m}\left[{t}\right]}$ and the decision variable ${[{\boldsymbol{\xi}_{k, m}}]_{i \coloneqq a_{k,m}\left[{t}\right]}}$ satisfying \eqref{eq_erank4pa_cons1}.
        			Then, the joint action ${\mathbf{a}^{\left({k}\right)}\left[{t}\right]}$ is  denoted as ${				\mathbf{a}^{\left({k}\right)}\left[{t}\right] = \left[{a_{k,1}\left[{t}\right], a_{k,2}\left[{t}\right], \cdots, a_{k,M_{\text{pa}}}\left[{t}\right]}\right]}$.
			
        		\subsubsection{Reward Design}
                    After all GRL agents have executed actions according to their observations at time slot ${t}$, the reward function ${\mathcal{R}_{\text{pa}}}$ will evaluate the quality of selected PA positions in terms of the optimization objective \eqref{eq_erank4pa_obj}, i.e., effective rank  ${{\mathbf{H}}_{\text{pa}}\left[{t}\right]}$ based on \eqref{eq_channel4pa}.
                    Therefore, the immediate reward ${r\left[{t}\right]}$in the built MAGRL framework is set as
                    \begin{equation}
                        \begin{aligned}
                        r\left[{t}\right] &= \mathcal{R}_{\text{pa}}({S_t, \{{\mathbf{a}^{\left({1}\right)}\left[{t}\right],
                        \mathbf{a}^{\left({2}\right)}\left[{t}\right],
                        \cdots, 
                        \mathbf{a}^{\left({K_{\text{wav}}}\right)}\left[{t}\right]}\}})\\
                        &= \rm{erank}\left({{\mathbf{H}}_{\text{pa}}\left[{t}\right]}\right) + \mathcal{C}_{\text{pa}}\left[t\right],
                        \end{aligned}
                    \end{equation}
                    where  ${{\mathcal{C}}_{\text{pa}}\left[ t \right] =  - \alpha \sum\nolimits_{k = 1}^{{K_{{\rm{pa}}}}} {\sum\nolimits_{m = 1}^{{M_{{\rm{pa}}}}} {\sum\nolimits_{m' = m + 1}^{{M_{{\rm{pa}}}}} {{\bf{\xi }}_{k,m}^{\top}{{\bf{\xi }}_{k,m'}}} } }  \leq 0}$ is a penalty to meet \eqref{eq_erank4pa_cons2}, based on the number of  collisions between PAs on the same waveguide.

        	\subsection{MAGQN Algorithm}
                To enable all GRL agents effectively learning their policy, a MAGAQN-based top-k PA positioning algorithm (labeled as the MAGAQN algorithm) is proposed to enhance the effective rank and ensure collision-free between multiple PAs on the same waveguide, which consists of the following three components: 
                \subsubsection{Graph Attention  Q-Network Structure}
        			As shown in \reffig{fig_scheme4magrl}, the graph attention \cite{velivckovic2017graph} is adopted to achieve the graph embedding of the input graph observation ${\mathbf{o}^{\left({k}\right)}}$, i.e., ${\mathbb{R}^{3+M_{\text{pa}}} \rightarrow \mathbb{R}^{d_{\text{emb}}}}$.
                    The graph embedding vector ${\mathbf{z}^{\left({k}\right)}}$ of the ${k}$-th GRL agent also is computed through both graph attention and graph pool layers, similar to \refsec{sec_algo4gaiqn}.
                    Meanwhile, to predict future mobility status by leveraging historical observations \cite{9839534}, the gated recurrent unit (GRU) is used to further handle the ${\mathbf{z}^{\left({k}\right)}}$, which is mathematically formulated by \cite{chung2014empirical}
        			\begin{subequations}
                        \begin{align}
                            \mathbf{r}^{\text{gru}}_{t} &= \rm{sigmoid}({\mathbf{z}^{\left({k}\right)}\left[{t}\right]\mathbf{w}_{\text{r1}} + \mathbf{h}^{\text{gru}}_{t}\mathbf{w}_{\text{r2}} + \mathbf{b}_{\text{r}}}),\\
                            \mathbf{z}^{\text{gru}}_{t}&=  \rm{sigmoid}({\mathbf{z}^{\left({k}\right)}\left[{t}\right]\mathbf{w}_{\text{z1}} + \mathbf{h}^{\text{gru}}_{t}\mathbf{w}_{\text{z2}} + \mathbf{b}_{\text{z}}}),\\
                            \tilde{\mathbf{h}}^{\text{gru}}_{t} &=\tanh({\mathbf{z}^{\left({k}\right)}\left[{t}\right]\mathbf{w}_{\text{h1}} +
                            \left({\mathbf{r}^{\text{gru}}_{t} \odot \mathbf{h}^{\text{gru}}_{t-1}}\right)\mathbf{w}_{\text{h2}}+
                            \mathbf{b}_{\text{h}}}),\\
                            \mathbf{h}^{\text{gru}}_{t} &=  \mathbf{z}^{\text{gru}}_{t} \odot \mathbf{h}^{\text{gru}}_{t-1} + \left({1-\mathbf{z}^{\text{gru}}_{t}}\right) \odot \tilde{\mathbf{h}}^{\text{gru}}_{t},
                        \end{align}
        			\end{subequations}where ${\mathbf{r}^{\text{gru}}_{t}}$ and ${\mathbf{z}^{\text{gru}}_{t}}$ are reset gate and update gate, respectively.
                    ${\mathbf{h}^{\text{gru}}_{t-1}}$ is the hidden state of the previous time step.
                    $\mathbf{w}_{\text{r1}}$, $\mathbf{w}_{\text{r2}}$, $\mathbf{w}_{\text{z1}}$, $\mathbf{w}_{\text{z2}}$, ${\mathbf{w}_{\text{h1}}}$ and  $\mathbf{w}_{\text{h2}}$ are weight parameters.
                    ${\mathbf{b}_{\text{r}}}$, ${\mathbf{b}_{\text{z}}}$ and ${\mathbf{b}_{\text{h}}}$ are biases.
                    ${\rm{sigmoid}}$ and ${\tanh}$ are nonlinear activation functions.
                    Then, its dueling structure can be expressed in \eqref{eq_dueling4gaqn}, where ${\boldsymbol{w}_{k}}$ indicates network parameters of the ${k}$-th agent.
                    \begin{figure*}[!ht]
                        \begin{equation}
                            Q_{\boldsymbol{w}_{k}}( {\mathbf{o}^{\left({k}\right)}\left[{t}\right], \mathbf{a}^{\left({k}\right)}\left[{t}\right]} ) = V^{\left({k}\right)}( {\mathbf{o}^{\left({k}\right)}\left[{t}\right]} ) + ( {A^{\left({k}\right)}( {\mathbf{o}^{\left({k}\right)}\left[{t}\right], \mathbf{a}^{\left({k}\right)}\left[{t}\right]} ) - \frac{1}{{\left| {{I_{pos}}} \right|}}\sum\nolimits_{a'} {A^{\left({k}\right)}( { \mathbf{o}^{\left({k}\right)}\left[ {t } \right], a'} )} } ).
                            \label{eq_dueling4gaqn}
                        \end{equation}
                        \rule{\textwidth}{0.4pt} 
                    \end{figure*}

                    \begin{figure*}[!ht]
                        \begin{equation}
                            \tag{27}
                            \mathscr{L}_i({\boldsymbol{w}_{k}}) = {r_i}\left[ t \right] + \lambda Q_{\boldsymbol{w}'_{k}}( {{\bf{o}}_i^{\left( k \right)}\left[ {t + 1} \right],\mathop {\rm{argtopk} }\nolimits_{{\bf{a}}_i^{\left( k \right)}\left[ {t + 1} \right]} Q_{\boldsymbol{w}_{k}}( {{\bf{o}}_i^{\left( k \right)}\left[ {t + 1} \right],{\bf{a}}_i^{\left( k \right)}\left[ {t + 1} \right]} )} ) - Q_{\boldsymbol{w}_{k}}( {\mathbf{o}_i^{\left( k \right)}\left[ t \right],\mathbf{a}_i^{\left( k \right)}\left[ t \right]} ).
                            \label{eq_loss4pa}
                        \end{equation}
                    \rule{\textwidth}{0.4pt} 
                    \end{figure*}

                    \begin{algorithm}[!t]
                        \caption{MAGAQN algorithm.}
                        \label{algo_MAGAQN}
                        \begin{algorithmic}[1]
                            \STATE Start the simulation system for the PA-enabled wireless system with ${K_{\text{wav}}}$ waveguides, ${M_{\text{pa}}}$ PAs and ${N}$ users;
                            \STATE Initialize the parameters ${\boldsymbol{w}_{k}}$ randomly;
                            \STATE Initialize the parameters ${\boldsymbol{w}'_{k}}$: ${\boldsymbol{w}'_{k}  \leftarrow \boldsymbol{w}_{k}}$;
                            \STATE \textbf{for} \textit{episode\_idx}$ = \left\{ {1,2, \cdots ,N_{\text{episode}}} \right\}$ \textbf{do}
                            \STATE \hspace{0.3cm} Reset positions of waveguides, PAs and users.
                            \STATE \hspace{0.3cm} \textbf{for} time slot $ t = \left\{ {1,2, \cdots , T} \right\}$ \textbf{do}
                            \STATE \hspace{0.3cm}\hspace{0.3cm} \textbf{for} each waveguide agent $ k = \left\{ {1,2, \cdots , K_{\text{wav}}} \right\}$ \textbf{do}
                            \STATE \hspace{0.3cm}\hspace{0.3cm}\hspace{0.3cm} 
                            \begin{minipage}[t]{.85\linewidth}
                            Observe ${\mathbf{o}^{\left({k}\right)}\left[{t}\right]}$ and then choose an action ${\mathbf{a}^{\left({k}\right)}\left[{t}\right]}$ based on \eqref{eq_action_selection4pa};
                            \end{minipage}
                            \STATE \hspace{0.3cm}\hspace{0.3cm} \textbf{end} \textbf{for}
                            \STATE \hspace{0.3cm}\hspace{0.3cm} All agents execute actions and receive reward ${r\left[t\right]}$;
                            \STATE \hspace{0.3cm}\hspace{0.3cm} \textbf{for} each waveguide agent $ k = \left\{ {1,2, \cdots , K_{\text{wav}}} \right\}$ \textbf{do}
                            \STATE \hspace{0.3cm}\hspace{0.3cm}\hspace{0.3cm} 
                            \begin{minipage}[t]{.87\linewidth}
                            Observe ${\mathbf{o}^{\left({k}\right)}\left[{t+1}\right]}$ and store the transition ($\mathbf{o}^{\left({k}\right)}\left[{t}\right]$,$ \mathbf{a}^{\left({k}\right)}\left[{t}\right]$,$ r\left[{t}\right]$, $\mathbf{o}^{\left({k}\right)}\left[{t+1}\right]$) into ${\mathbb{D}_{k}}$.
                            \end{minipage}
                            \STATE \hspace{0.3cm}\hspace{0.3cm} \textbf{end} \textbf{for}
                            \STATE \hspace{0.3cm} \textbf{end} \textbf{for}	
                            \STATE \hspace{0.3cm} \textbf{for} each waveguide agent $ k = \left\{ {1,2, \cdots , K_{\text{wav}}} \right\}$ \textbf{do}
                            \STATE \hspace{0.3cm}\hspace{0.3cm}
                            \begin{minipage}[t]{.9\linewidth}
                            Sample a mini-batch of size ${I_{\text{batch}}}$ from ${\mathbb{D}_{k}}$ ;
                            \end{minipage}
                            \STATE \hspace{0.3cm}\hspace{0.3cm}
                            \begin{minipage}[t]{.9\linewidth}
                            For each transition ${i \in \left\{{1, 2, \cdots, I_{\text{batch}}}\right\}}$, calculate  ${\hat{\mathscr{L}}_i\left({\boldsymbol{w}_k}\right)}$ based on \eqref{eq_per_loss4pa} to update ${\boldsymbol{w}_k}$ .
                            \end{minipage}
                            \STATE \hspace{0.3cm}\hspace{0.3cm}
                            \begin{minipage}[t]{.9\linewidth}
                            Soft update ${\boldsymbol{w}'_k}$ based on \eqref{eq_soft_update4pa};
                            \end{minipage}
                            \STATE \hspace{0.3cm} \textbf{end} \textbf{for}
                            \STATE \textbf{end} \textbf{for}
                            \STATE \textbf{return:} the parameters ${\boldsymbol{w}_k}$ of each main network.
                        \end{algorithmic}
                    \end{algorithm}

                \subsubsection{Top-k Action Selection for PA system}
                    To effectively avoid multiple PAs from simultaneously activating the same position, i.e., satisfying constraint \eqref{eq_erank4pa_cons2}, the top-k action selection method in \refsec{sec_topk} is also adopted for each waveguide as GRL agent. 
                    Hence, at each time slot ${t}$, each GRL agent can sample ${M_{\text{pa}}}$ non-repeating actions with a probability of ${\varepsilon \left( t \right)}$ and exploit the action ${\mathbf{a}^{\left({k}\right)}\left[{t}\right]}$, with the probability ${1 - \varepsilon \left( t \right)}$, i.e.,
                    \begin{equation}
                        \mathbf{a}^{\left({k}\right)}\left[{t}\right] = 
                        \begin{cases}
                        \arg {\rm{top}}{{\rm{k}}_{\mathbf{a}}}\left( {{Q_{\boldsymbol{w}_{k}}}\left( {\mathbf{o}^{\left({k}\right)}\left[ t \right], \mathbf{a}} \right), M_{\text{pa}}} \right), {\text{if }} 1 - \varepsilon \left( t \right),\\ 
                        {\text{random},}{\text{otherwise.}} 
                    \end{cases}
                    \label{eq_action_selection4pa}
                    \end{equation}

        		\subsubsection{Parameter Update}
                    For the ${k}$-th each GRL agent , its loss function of the ${i}$-th transition at time slot ${t}$ can be defined in \eqref{eq_loss4pa}.
                    Since the PER buffer ${\mathbb{D}_{k}}$ is also utilized, its loss function of prioritized weight can be expressed by
                    \setcounter{equation}{27}
                    \begin{equation}
                        \hat{\mathscr{L}}_i\left({\boldsymbol{w}_{k}}\right) = \left({\frac{1}{\left|{\mathbb{D}_{k}}\right|}  \cdot \frac{1}{P\left({i}\right)}}\right)^{\beta} \mathscr{L}_i\left({\boldsymbol{w}_{k}}\right).
                        \label{eq_per_loss4pa}
                    \end{equation}
                    Furthermore, its target network uses the soft update method, parameterized by ${\tau_{\text{soft}} \in \left({0, 1}\right)}$, which is defined by
                    \begin{equation}
                        \boldsymbol{w}'_{k} \leftarrow {\tau _{{\rm{soft}}}}\boldsymbol{w}_{k} + \left( {1 - {\tau _{{\rm{soft}}}}} \right)\boldsymbol{w}'_{k}.
                        \label{eq_soft_update4pa}
                    \end{equation}

        			In summary, the workflow of the MAGAQN algorithm is summarized in \refalg{algo_MAGAQN}.

            \subsection{Complexity Analysis}
            \subsubsection{Training Stage} 
            The computational complexity of the proposed MAGAQN algorithm mainly involves four parts.
            \begin{itemize}
            \item \textbf{Network Initialize:}
            This phase involves the initialization of network parameters for graph attention \cite{velivckovic2017graph}, GRU \cite{9839534} and dueling architecture components of ${K_{\text{wav}}}$ GRL agents, i.e., the computational complexity is represented by $\mathcal{O}({\sum\nolimits_{k = 1}^{{K_{{\rm{wav}}}}} {\left| {{\boldsymbol{w}_k}} \right|}  + \left| {{{\boldsymbol{w}'}_k}} \right|})$, where ${\mathcal{O}\left({\left| {{\boldsymbol{w}'_k}} \right|}\right)=\mathcal{O}({\left| {{\boldsymbol{w}_k}} \right|})}$ and is expressed as ${\mathcal{O}(\sum\nolimits_{l = 1}^{{L_{{\rm{emb}}}}} d_{{\rm{in}}}^{\left( l \right)} d_{{\rm{out}}}^{\left( l \right)}  + {F_{{\rm{hid}}}}( {d_{{\rm{emb}}}} +8}$
            ${{F_{{\rm{hid}}}} + 1 )}$
            ${ + \sum\nolimits_{l = 1}^{{L_{{\rm{hid}}}}} {F_{{\rm{in}}}^{\left( l \right)}F_{{\rm{out}}}^{\left( l \right)}} )}$.
            
            \item \textbf{Action Sampling:} Actions are generated based on the current graph observation of each GRL agent, and the computational complexity is given by
            ${\mathcal{O}({N_{\text{episode}} T \sum\nolimits_{k = 1}^{{K_{{\rm{wav}}}}} {\left| {{{\hat{\boldsymbol{w}}}_k}} \right|} })}$, where ${\mathcal{O}(\left| {{{\hat{\boldsymbol{w}}}_k}} \right|)}$ is the computational complexity of network inference. 
            It can be expressed as 
            ${\mathcal{O} ( \sum\nolimits_{l = 1}^{{L_{{\rm{emb}}}}}( \left| {\mathcal{N}_k^{{\rm{wav}}}} \right|d_{{\rm{in}}}^{\left( l \right)} + \mathcal{E}_{{\rm{pa}}}^k )d_{{\rm{out}}}^{\left( l \right)}+}$
            ${\left| {\mathcal{N}_k^{{\rm{wav}}}} \right|{d_{{\rm{emb}}}}+ F_{\text{hid}}\left( {{d_{{\rm{emb}}}} + 8{F_{{\rm{hid}}}} + 1} \right)
            +\sum\nolimits_{l = 1}^{{L_{{\rm{hid}}}}} {F_{{\rm{in}}}^{\left( l \right)}F_{{\rm{out}}}^{\left( l \right)}}}$
            ${  + {F_{{\rm{hid}}}}\left( {{I_{{\rm{pos}}}} + 1} \right)+ {I_{{\rm{pos}}}})}$.
            
            \item \textbf{PER Buffer Collection:} The complexity of collecting transitions in the PER buffer is ${\mathcal{O}\left({N_{\text{episode}}K_{\text{wav}} \left({TV + I_{\text{batch}}}\right)}\right)}$.
            \item \textbf{Network Update:} Its computational complexity can be computed by
            ${\mathcal{O}({2 N_{\text{episode}} \sum\nolimits_{k = 1}^{{K_{{\rm{wav}}}}} {\left| {{{\boldsymbol{w}}_k}} \right|} })}$. 
            \end{itemize}

        In the training stage, the space complexity of the MAGAQN is 
        $\mathcal{O}(\sum\nolimits_{k = 1}^{{K_{{\rm{wav}}}}}(
        2\left| {{{\boldsymbol{w}}_k}} \right|
        +\left|{\mathbb{D}_{k}}\right|( 
            2\left|{\mathcal{N}_{k}^{\text{wav}}}\right|( 
                3+K_{\text{wav}}
                +\left|{\mathcal{N}_{k}^{\text{wav}}}\right|
            )
            + M_{\text{pa}} + 1
        )
        )),$
        where ${\left|{\mathbb{D}_k}\right|}$ is the size of the PER buffer for each GRL agent.

		\subsubsection{Execution Stage}
		The computational complexity of the MAGAQN is ${\mathcal{O}({\sum\nolimits_{k = 1}^{{K_{{\rm{wav}}}}} {\left| {{\hat{\boldsymbol{w}}_k}} \right|} })}$, which is primarily due to the main network inferring actions based on the current graph observation.
		Moreover, the space complexity in this phase can be expressed by ${\mathcal{O}({\sum\nolimits_{k = 1}^{{K_{{\rm{wav}}}}} {\left| {{\boldsymbol{w}_k}} \right|} })}$.

	\section{Simulation Results \label{sec_simulation}}

        This section presents extensive simulations to evaluate the performance of our proposed \refalg{algo_GAIQN} and \refalg{algo_MAGAQN} in terms of improving the effective rank of MA and PA-enabled wireless systems over multi-time slots.
		\subsection{Experiment Settings}
        
        			This subsection details the simulation setup, including the simulation platform, simulation setup, and benchmarks adopted to evaluate the performance of our proposed GAIQN and MAGAQN algorithms.
        
           		\subsubsection{Simulation Platform}
           			All experiments were conducted on a workstation equipped with an NVIDIA GeForce RTX 5080 GPU (16 GB) and an AMD Ryzen 9 9950X3D 16-core CPU running at 4.3 GHz, with 32 GB of RAM. The system operates on Ubuntu 24.04.3 LTS. 
           			The simulation environments of MA and PA systems are implemented in Python 3.11.9, with NetworkX 3.6.1 used for user graph construction and PyTorch 2.9.1+cu129 for learning-based computations. 
           			Additionally, torch-geometric 2.7.0 and cuGraph 25.12.02 are employed to implement and accelerate graph embedding computations, respectively.
                
        	\subsubsection{Simulation Setup}
        		In the simulations, we assume that each user's position at each time slot follows the uniform distribution, i.e., ${\forall x_{n}\left[{t}\right], y_{n}\left[{t}\right] \sim \mathcal{U}\left[{0, D_{\text{area}}}\right]}$.
                In the MA system, a geometric channel model is assumed with identical numbers of  transmit and receive paths for all users \cite{10304448}, i.e., ${L^{\text{t}}=L^{\text{r}}=L}$.
        		Hence, each diagonal element of PRM ${\mathbf{\Sigma}_n}$ for each user is an i.i.d. CSCG random variable, expressed by ${\forall g_{n,l} \sim \mathcal{CN}({0, \rho d_{n, \text{BS}}^{-\varsigma}/L})}$, where ${\rho d_{n, \text{BS}}^{-\varsigma}}$ denotes the expected channel gain with ${\varsigma }$ indicating the path loss exponent and ${\rho}$ representing the path loss at a reference distance of 1 meter (m) \cite{10741192}.
                For each user, the azimuth and elevation AoAs and AoDs are assumed to follow the i.i.d. uniform distribution ${\mathcal{U}\left[{-\pi/2, \pi/2}\right]}$.
        		For the PA system, the waveguide length is assumed to be equal to the length of the square area \cite{11223640, 11267231}.
        		Then, the parameters of simulations and proposed algorithms are detailed in \reftable{table_sim_param} and \reftable{table_param4algo}, respectively.

            \begin{table}[!t]
                \caption{Simulation Parameters}
                \label{table_sim_param} 
                \centering
                \begin{tabular}{c| l| l}
                    \toprule
                    \textbf{System}  &\textbf{Parameter} &\textbf{Symbol/Value}\\
                    \midrule
                    \multirow{4}{*}{MA/PA} 
                    &Number of time slots & ${T=10}$\\
                    \cline{2-3}
                    &Carrier wavelength \cite{10741192} &${\lambda = 0.1}$ (m)\\
                    \cline{2-3}
                    &Antenna height of user \cite{3gpp.38.901} & ${z_{n} = 1.5}$ (m)\\
                    \cline{2-3}
                    &\begin{minipage}[t]{.55\linewidth}Number of antenna candidate positions\end{minipage} & ${I_{\text{pos}=100}}$\\
                    \hline
                    \multirow{5}{*}{MA}
                    &\begin{minipage}[t]{.55\linewidth}Number of MAs \cite{10741192}\end{minipage} 
                    &${M_{\text{ma}}=16}$ \\
                    \cline{2-3}
                    &\begin{minipage}[t]{.55\linewidth} 
                        Number of channel paths for each user \cite{10741192}
                    \end{minipage}
                    &${L=10}$\\
                    \cline{2-3}
                    &\begin{minipage}[t]{.55\linewidth}Channel gain at the reference distance \cite{10741192}\end{minipage} &${\rho=-40}$ (dB)\\
                    \cline{2-3}
                    &Path loss exponent \cite{10741192} &${\varsigma=2.8}$\\
                    \cline{2-3}
                    &\begin{minipage}[t]{.55\linewidth}Minimum inter-MA distance \cite{10741192}\end{minipage}  & ${D=\lambda/2}$  (m)\\
                    \cline{2-3}
                    &Antenna height of BS \cite{3gpp.38.901} & ${z_{\text{BS}} = 25}$ (m)\\
                    \hline
                    \multirow{3}{*}{PA}
                    &\begin{minipage}[t]{.55\linewidth}Number of PAs on each waveguide\end{minipage} &${M_{\text{pa}}=2}$\\
                    \cline{2-3}
                    &Number of waveguides \cite{11165763} &${K_{\text{wav}}=8}$\\
                    \cline{2-3}
                    &Effective refractive index \cite{11165763} &${n_{\text{eff}}=1.4}$\\
                    \cline{2-3}
                    &Height of each waveguide \cite{11165763} &${z_{\text{wav}}=3.0}$ (m)\\
                    \bottomrule
                \end{tabular}
            \end{table}
            
            \begin{table}[!t]
                \caption{Key Parameters of  Proposed Schemes}
                \label{table_param4algo} 
                \centering
                \begin{tabular}{l | c c}
                    \toprule
                    \multirow{2}{*}{\textbf{Parameters}}               &\multicolumn{2}{c}{\textbf{Value}}                \\
                    \cline{2-3}
                    &GRL  &MAGRL               \\
                    \midrule
                    the capacity of replay buffer   		  &\multicolumn{2}{c}{5000}                    \\
                    size of mini-batch                     	  &\multicolumn{2}{c}{64}                      \\
                    discount factor		                 	  &\multicolumn{2}{c}{0.98}                    \\
                    
                    soft update coefficient                	  &\multicolumn{2}{c}{0.005}                  \\
                    neurons' number of each embedding layer   &\multicolumn{2}{c}{256}                     \\
                    neurons' number of each hidden layer      &\multicolumn{2}{c}{256}                     \\
                    
                    initial exploration factor                &\multicolumn{2}{c}{1.0}                    \\
                    end exploration factor                    &\multicolumn{2}{c}{0.05}                     \\
                    anneal time of exploration factor	      &\multicolumn{2}{c}{10000}                    \\
                    the hyperparameter of penalty &\multicolumn{2}{c}{0.5}\\
                    the threshold of distance &\multicolumn{2}{c}{20 (m)} \\
                    the threshold of angle &\multicolumn{2}{c}{0.3  (rad)}\\
                    
                    \hline
                    threshold of Huber loss                	  &1.0     &${\backslash}$                \\
                    number of value distribution              &64	 &${\backslash}$ 				   \\
                    number of cosines                         &32		 &${\backslash}$ 			   \\
                    learning rate            				  &0.01   &0.001                 \\
                    descent method of exploration factor      &"liner" &"exp"                \\
                    \bottomrule
                \end{tabular}        
            \end{table}

        	\subsubsection{Benchmarks}

            	To validate the effectiveness of our proposed  GAIQN algorithm in the MA-enabled multi-user downlink communication, we utilize the QMIX-ATT \cite{yu2023enhancing}, QATTEN  \cite{yang2020qatten} and random algorithms as benchmarks, as follows:
            	\begin{itemize}
            
            		\item \textbf{QMIX-ATT} \cite{yu2023enhancing}:
                   Each MA is modeled as an independent agent equipped with a Q-network to select its position in the 2D array panel.
            		QMIX-ATT is an attention-enhanced variant of the QMIX framework, designed to improve coordination by explicitly modeling inter-agent dependencies during centralized training.
            		Compared with the original QMIX, an attention mechanism is incorporated to adaptively emphasize relevant agent information and contextual features, enabling more effective joint action-value estimation in dynamic multi-agent environments.
            		During training, a centralized mixing network leverages the global state information to jointly update the individual Q-networks of all agents while preserving decentralized execution at runtime.
            		To balance exploration and exploitation, each agent adopts a ${\varepsilon}$-greedy policy.
            
            		\item \textbf{QATTEN} \cite{yang2020qatten}:
            		Each MA maintains an independent Q-network to estimate the action values associated with its positions in the 2D array panel.
            		An attention-based aggregation module is employed to combine the individual Q-values of all MAs into a joint action–value function, where attention weights are adaptively generated from agent representations and global contextual information.
            		Meanwhile, it enables the training process to dynamically emphasize more influential agents and capture inter-agent interactions.
            		The aggregated joint Q-value is then used to compute the training loss and jointly update the Q-networks of all MAs, facilitating effective coordination learning for cooperative position selection.
            		
            		\item \textbf{RANDOM}: At each time slot, the BS randomly generates positions for all MAs. 
            	    It is adopted as a baseline without any learning or optimization.
            	\end{itemize}
            
            	For comparison for our proposed MAGAQN algorithm in the PA-enabled multi-user downlink communication, we utilize the MAPPO \cite{yu2022surprising}, IPPO \cite{de2020independent} and random algorithms as benchmarks, as follows
            	\begin{itemize}
            		\item \textbf{MAPPO} \cite{yu2022surprising}: Each PA is modeled as an agent equipped with an independent policy network to select its candidate position. 
                   During the training phase, all agents share a centralized value network that exploits the global state and joint action information to accurately evaluate the joint state value . This centralized evaluation guides the policy updates, thereby ensuring consistent and coordinated learning among agents.
            		
            		\item \textbf{IPPO} \cite{de2020independent}: Each PA is modeled as an agent equipped with an independent policy and an independent value network. 
            		During training, the framework corresponds to a parallel implementation of independent proximal policy optimizations , where each agent updates its policy and value networks solely based on local observations, actions, and rewards, without any information sharing, effectively operating in a non-stationary single-agent environment.
            		
            		\item \textbf{RANDOM}: At each time slot, each waveguide randomly selects positions for all MAs. 
            		It is used as a baseline in the absence of any learning or optimization.
            	\end{itemize}
            

	\subsection{Convergence Performance Analysis}
        \begin{figure*}[!ht]
            \centering
            \subfloat[]{\includegraphics[width=0.33\linewidth]{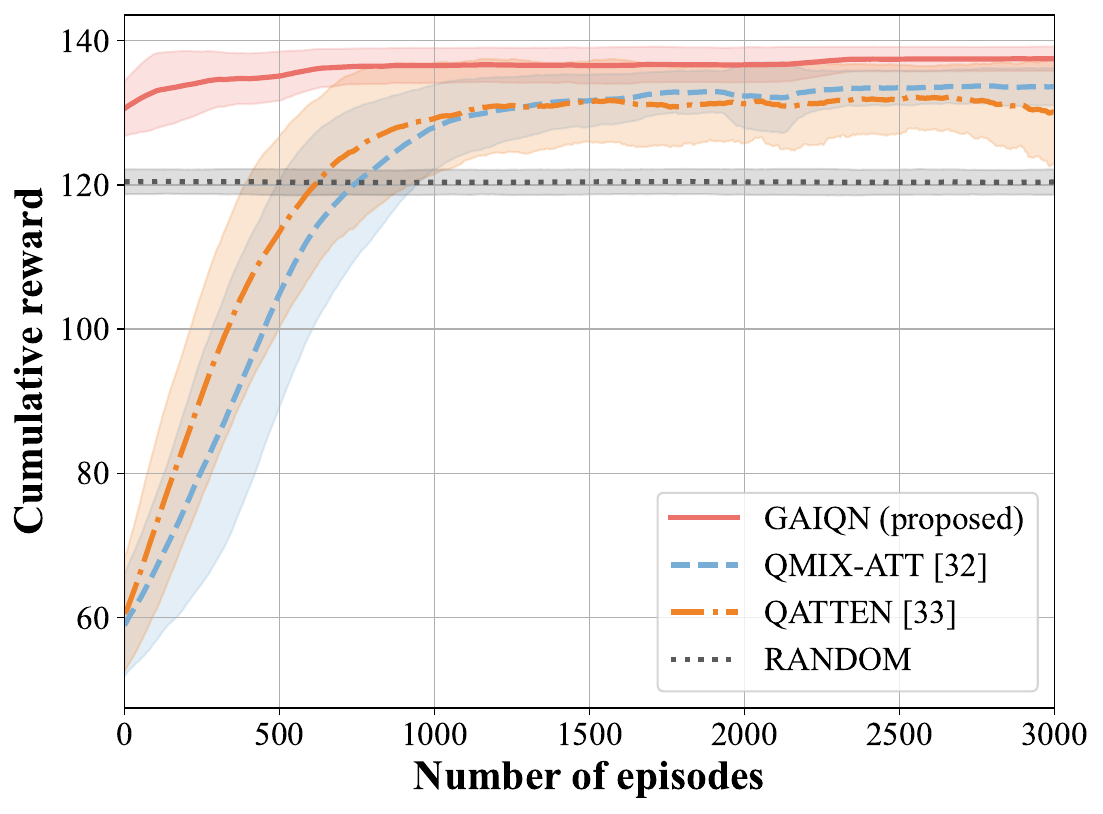}\label{fig_a_tpa_ma}}
            \hfil
            \subfloat[]{\includegraphics[width=0.33\linewidth]{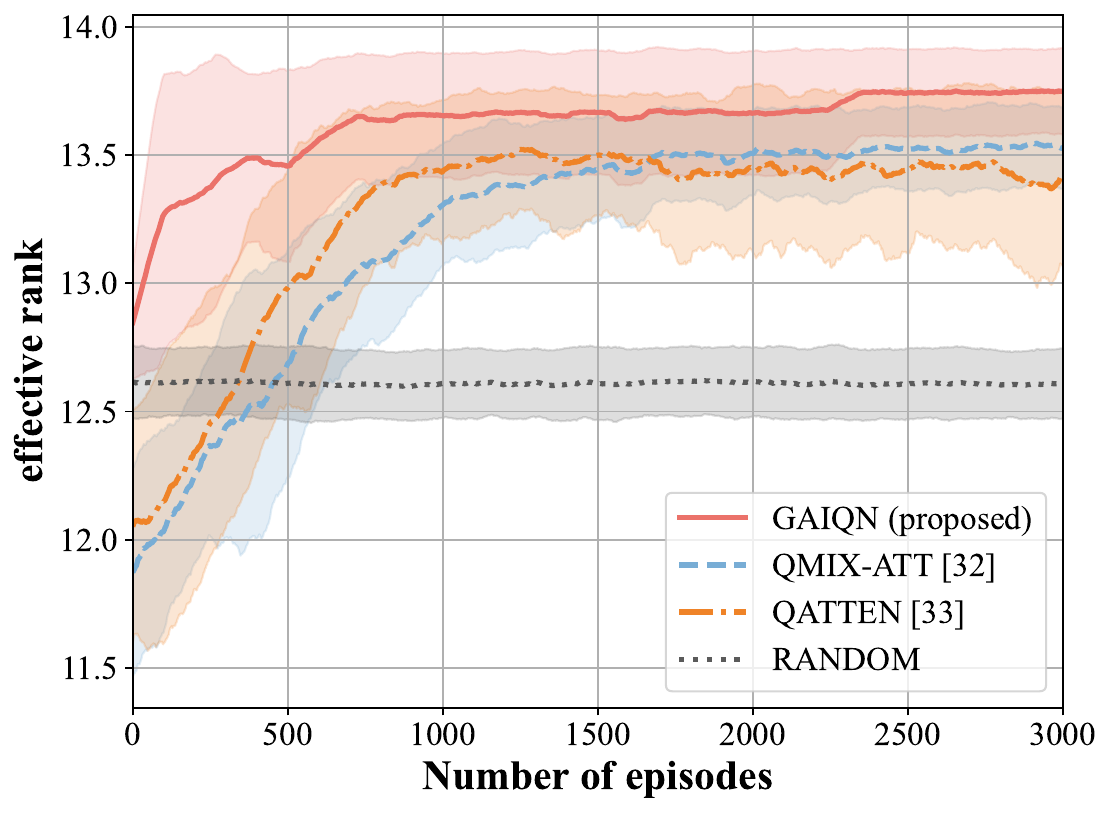}\label{fig_b_tpa_ma}}
            \hfil
            \subfloat[]{\includegraphics[width=0.33\linewidth]{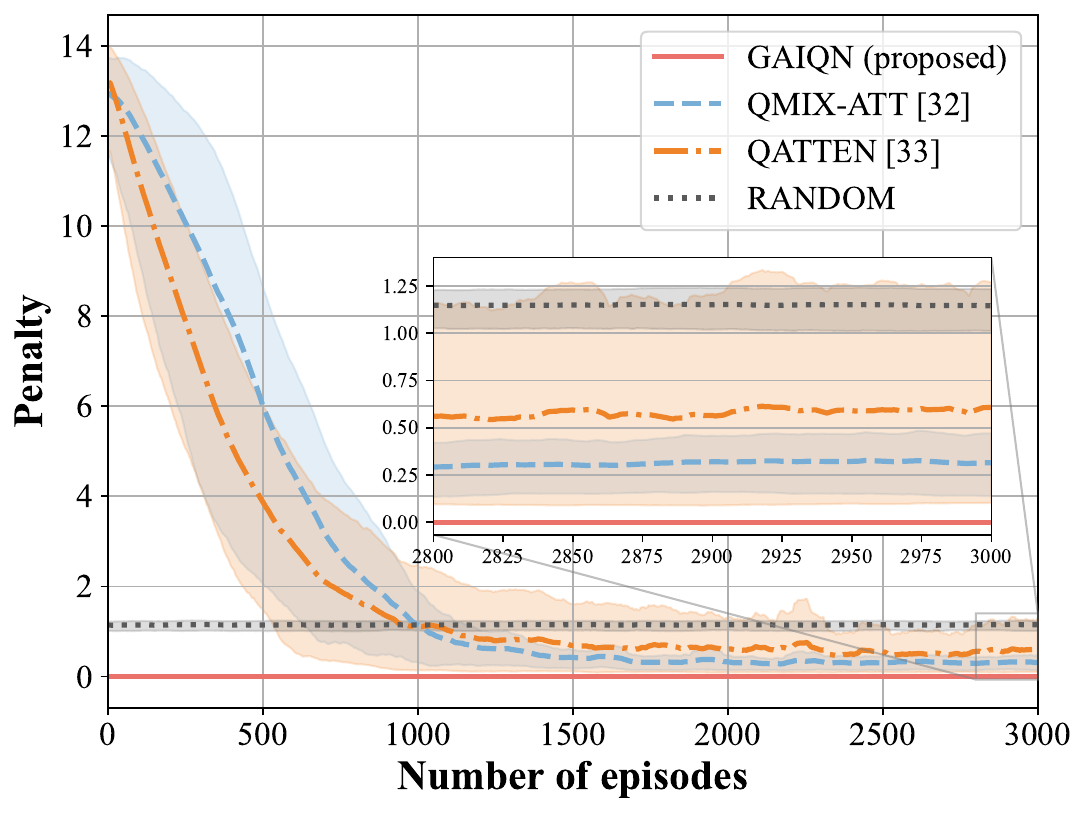}\label{fig_c_tpa_ma}}
            \caption{Convergence performance for the MA system under different algorithms. (a) Cumulative reward. (b) Effective rank. (c) Penalty.}
            \label{fig_tpa_ma}
                        \vspace{-0.5cm}
        \end{figure*}
        \begin{figure*}[!ht]
        \centering
        \subfloat[]{\includegraphics[width=0.33\linewidth]{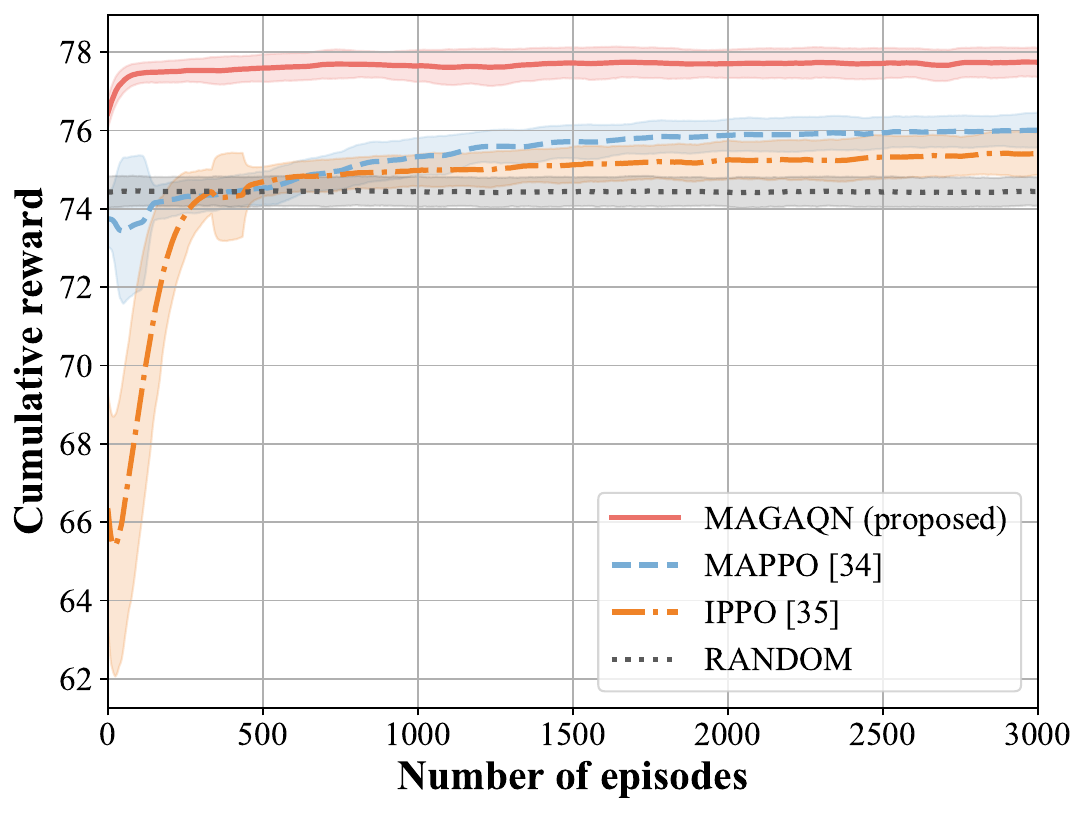}\label{fig_a_tpa_pa}}
        \hfil
        \subfloat[]{\includegraphics[width=0.33\linewidth]{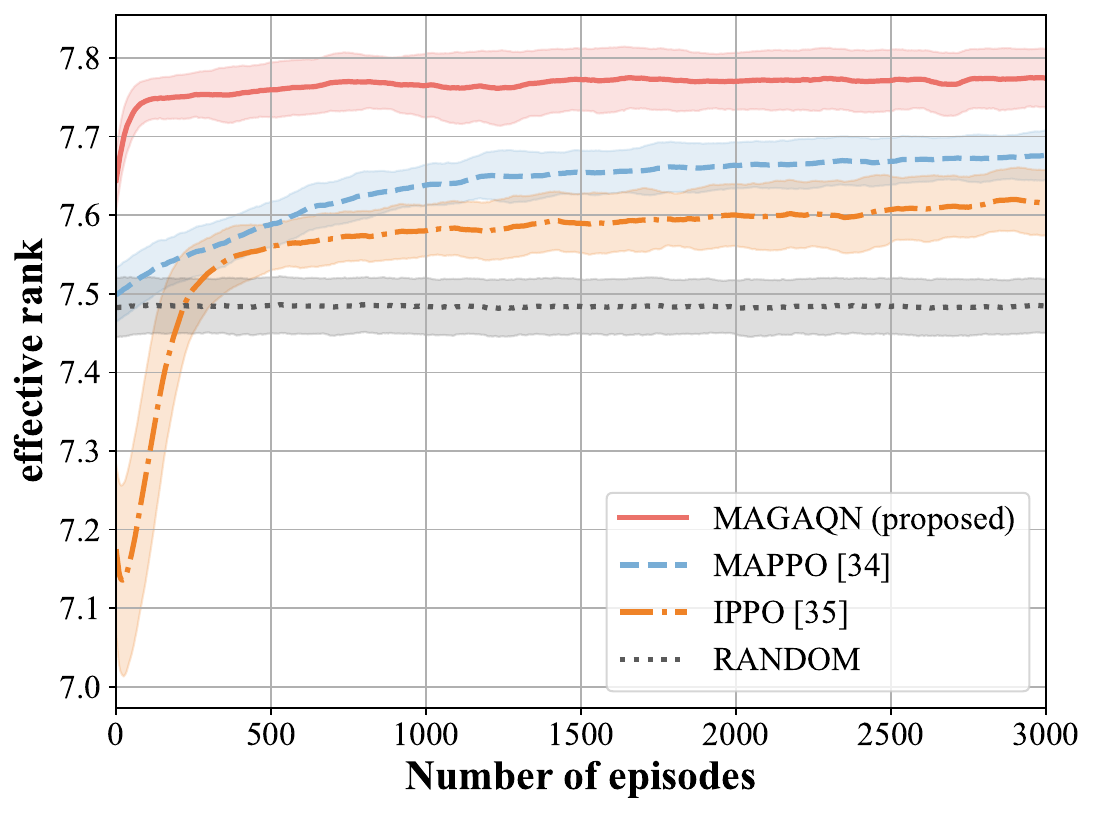}\label{fig_b_tpa_pa}}
        \hfil
        \subfloat[]{\includegraphics[width=0.33\linewidth]{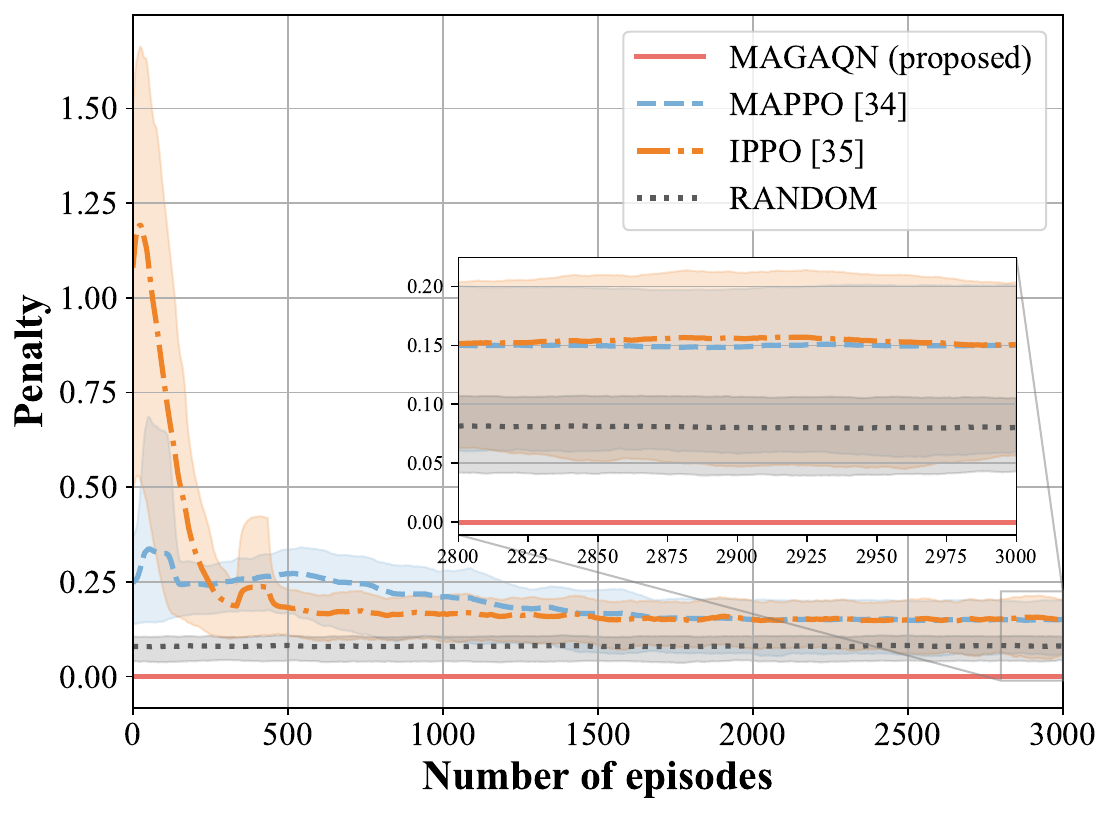}\label{fig_c_tpa_pa}}
        \caption{Convergence performance for the PA system under different algorithms. (a) Cumulative reward. (b) Effective rank. (c) Penalty.}
        \label{fig_tpa_pa}
        \end{figure*}

        We conduct a training within a ${200 \times 200 }$ ${\text{m}^{2}}$ square area with 80 users, considering both the MA system with 16 MAs and the PA system consisting of 8 waveguides, each equipped with 2 PAs.
		\reffig{fig_tpa_ma} illustrates the convergence performance of our proposed GAIQN algorithm and benchmarks in terms of cumulative reward, effective rank, and penalty.
        In \reffig{fig_tpa_ma}(a), the GAIQN achieves the highest cumulative reward and table convergence, while QMIX-ATT shows lower performance with higher variance compared to GAIQN.
		Then, QATTEN improves rapidly in the early training episode and converges faster than the QMIX-ATT algorithm, but its converged cumulative reward remains lower than that of QMIX-ATT and GAIQN.
		The superior performance of GAIQN over QMIX-ATT and QATTEN stems from the ability to effectively extract user spatial distribution information and directly avoid collisions between MAs.
		QMIX further outperforms QATTEN by incorporating self-attention directly into each agent’s Q-network, enabling dynamic awareness of the states and actions of other relevant agents during decision making. 
        In contrast, QATTEN applies self-attention at the mixer network to reweigh and aggregate the individual agents’ Q-values, without explicitly enhancing each agent’s local decision process.
		
		The convergence trend in \reffig{fig_tpa_ma}(b) demonstrates that the proposed GAIQN algorithm achieves a consistent increase in the effective rank, ultimately converging to the highest value among all algorithms. 
		GAIQN exhibits a faster convergence behavior compared to the QMIX-ATT and QATTEN, which can be primarily attributed to its top-k action selection method, which makes the GRL agent only focus on maximizing the effective rank, which is the actual performance metric of interest, rather than jointly handling collision between MAs and improving effective rank. 
		In contrast, QIMX-ATT and QATTEN mainly rely on the penalty-based reward function to avoid invalid actions, leading to a slower convergence.
        RANDOM maintains a consistently low effective rank throughout training with minimal variation, indicating the absence of learning capability.
        Overall, the GAIQN algorithm offers  a 1.6\% to 9.0\% performance improvement compared with benchmarks.

		\reffig{fig_tpa_ma}(c) shows that the penalties of different algorithms in MA systems.
        By incorporating self-attention into the Q-network, QMIX-ATT captures inter-agent dependencies and resolves action conflicts, thereby achieving lower penalties than QATTEN.
        However, although QMIX-ATT and QATTEN algorithms can reduce penalties, occasional collisions between multiple MAs still occur.
        For the GAIQN algorithm, its top-k action selection method explicitly prevents multiple MAs from selecting the same position on the 2D array plane.
        As a result, collision-free operation is guaranteed throughout training, and the penalty remains consistently zero.

		\reffig{fig_tpa_pa} illustrates the convergence results of our proposed MAGAQN algorithm and benchmarks.
		In \reffig{fig_tpa_pa}(a), the MAGAQN achieves the highest cumulative reward and fastest convergence compared with benchmarks.
		MAGAQN explicitly exploits user spatial correlations through graph attention, embedding the user spatial topology into graph embedding vectors. Since geographically proximate users exhibit correlated channel states, this graph-based encoding effectively captures spatial interference dependencies. 
		In contrast, MAPPO lacks explicit spatial reasoning, resulting in inferior multi-agent coordination and lower cumulative reward.
		MAPPO employs a centralized value function during training to mitigate non-stationarity and coordinate agents, yielding higher, more stable asymptotic performance than IPPO. 

\begin{table*}[!t]
    \caption{Effective Rank Versus Various System Parameters}
    \label{table_eRank_as_various_system_parameters} 
    \centering
    \begin{tabular}{c|c|c|c|c|c|c}
        \hline
        \multicolumn{7}{c}{\textbf{Effective rank versus number of users}}\\
        \hline
        & Algorithm &${N=40}$  &${N=50}$ &${N=60}$ &${N=70}$ &${N=80}$\\
        \hline
        \multirow{4}{*}{MA system}
        &GAIQN (proposed) &$\mathbf{12.555\pm0.098}$     &$\mathbf{12.986\pm0.107}$     &$\mathbf{13.290\pm0.106}$     &$\mathbf{13.530\pm0.109}$     &$\mathbf{13.721\pm0.106}$ \\
        &QMIX-ATT \cite{yu2023enhancing}&$12.416\pm0.077$     &$12.825\pm0.093$     &$13.122\pm0.103$     &$13.354\pm0.102$     &$13.538\pm0.098$ \\
        &QATTEN \cite{yang2020qatten}&$12.266\pm0.115$     &$12.677\pm0.117$     &$12.968\pm0.125$     &$13.199\pm0.126$     &$13.382\pm0.122$ \\
        &RANDOM &$11.622\pm0.070$     &$11.982\pm0.086$     &$12.243\pm0.090$     &$12.452\pm0.088$     &$12.617\pm0.081$ \\
        \hline
        \multirow{4}{*}{PA system}
        &MAGAQN (proposed) &$\mathbf{7.541\pm0.021}$     &$\mathbf{7.631\pm0.021}$     &$\mathbf{7.693\pm0.021}$     &$\mathbf{7.740\pm0.022}$     &$\mathbf{7.777\pm0.023}$ \\
        &MAPPO \cite{yu2022surprising} &$7.442\pm0.016$     &$7.534\pm0.015$     &$7.595\pm0.015$     &$7.643\pm0.014$     &$7.681\pm0.015$ \\
        &IPPO \cite{de2020independent} &$7.374\pm0.025$     &$7.463\pm0.026$     &$7.528\pm0.026$     &$7.576\pm0.027$     &$7.615\pm0.028$ \\
        &RANDOM &$7.248\pm0.013$     &$7.333\pm0.012$     &$7.397\pm0.012$     &$7.443\pm0.010$     &$7.482\pm0.010$ \\
        \hline
        \hline
        \multicolumn{7}{c}{\textbf{Effective rank versus length of square area}}\\
        \hline
        &Algorithms  &${D_{\text{aera}}=120}$ m  &${D_{\text{aera}}=140}$ m &${D_{\text{aera}}=160}$ m &${D_{\text{aera}}=180}$ m &${D_{\text{aera}}=200}$ m\\
        \hline
        \multirow{4}{*}{MA system}
        &GAIQN (proposed) 
        &$\mathbf{14.387\pm0.042}$     &$\mathbf{14.166\pm0.080}$     &$\mathbf{14.021\pm0.047}$     &$\mathbf{13.877\pm0.057}$     &$\mathbf{13.721\pm0.106}$ \\
        &QMIX-ATT\cite{yu2023enhancing} &$14.223\pm0.023$     &$14.031\pm0.061$     &$13.877\pm0.085$     &$13.74\pm0.054$     &$13.538\pm0.098$ \\
        &QATTEN\cite{yang2020qatten} &$14.013\pm0.068$     &$13.779\pm0.085$     &$13.618\pm0.101$     &$13.563\pm0.117$     &$13.382\pm0.122$ \\
        &RANDOM &$13.222\pm0.018$     &$13.065\pm0.046$     &$12.936\pm0.038$     &$12.804\pm0.04$     &$12.617\pm0.081$ \\
        \hline
        \multirow{4}{*}{PA system}
        &MAGAQN (proposed) 
        &$\mathbf{7.789\pm0.006}$     &$\mathbf{7.780\pm0.005}$     &$\mathbf{7.784\pm0.010}$     &$\mathbf{7.775\pm0.016}$     &$\mathbf{7.777\pm0.023}$ \\         
        &MAPPO\cite{yu2022surprising} &$7.732\pm0.009$     &$7.715\pm0.010$     &$7.716\pm0.005$     &$7.690\pm0.011$     &$7.681\pm0.015$ \\
        &IPPO\cite{de2020independent} &$7.646\pm0.023$     &$7.627\pm0.026$     &$7.673\pm0.032$     &$7.624\pm0.021$     &$7.615\pm0.028$ \\
        &RANDOM &$7.578\pm0.008$     &$7.547\pm0.010$     &$7.523\pm0.011$     &$7.496\pm0.006$     &$7.482\pm0.010$ \\
        \hline			
    \end{tabular}
\end{table*}

		As shown in \reffig{fig_tpa_pa}(b), the proposed MAGAQN algorithm consistently improves the effective rank throughout training and achieves the highest value compared to other algorithms.
		This confirms that the learning process of MAGAQN, MAPPO and IPPO leads to genuine performance enhancement rather than merely optimizing the cumulative reward. 
        RANDOM maintains a consistently low effective rank throughout training with minimal variation, indicating the absence of learning capability.
		Overall, the MAGAQN algorithm achieves a 1.3\% to 3.9\%  performance improvement compared with benchmarks.
        
        \reffig{fig_tpa_pa}(c) shows that the penalties of different algorithms in PA systems.
		Due to the adoption of the top-k action selection method, the Q-network in each GRL agent outputs the selection scores for all candidate positions, based on which the top-k positions are chosen.
		This unified decision process inherently prevents multiple PAs on the same waveguide from selecting the same candidate position, thereby eliminating antenna collisions.
		MAPPO and IPPO exhibit inferior performance and even underperform the RANDOM due to their inability to effectively handle PA collisions.
		In these algorithms, each PA operates with an independent action output, and coordination feasibility is not enforced at the action level, leading to multiple PAs occupying the same position. 

	\subsection{Effective Rank Versus Various System Parameters}

		\reftable{table_eRank_as_various_system_parameters} demonstrates the effective rank versus the number of users ${N}$ with all algorithms, considering the length of the square area ${D_{\text{area}}=200}$ m for MA and PA systems with 16 antennas.
		Increasing the number of users leads to a pronounced improvement in the effective rank of both MA and PA systems.
		This is because increasing the number of users enriches the spatial heterogeneity of the multi-user channel, leading to a more evenly distributed singular value spectrum and hence a higher effective rank.
		The PA-enabled wireless system is inherently constrained by its waveguide-based architecture, in which multiple PAs sharing the same waveguide contribute coherently and are effectively aggregated, thereby reducing the number of independent spatial DoF and leading to a lower-dimensional channel matrix with a smaller effective rank. 
		In contrast, MA elements are independently reconfigurable and preserve more independent channel dimensions. 
        Under a fair comparison between an MA system with 16 MAs and a PA system consisting of 8 waveguides, each with 2 PAs, the PA system consistently exhibits a lower effective rank  than the MA system due to the correlated and aggregated channel responses along each waveguide.
        Therefore, under the same number of antennas, the MA system improves the effective rank by 66.5\% to 76.4\% compared with the PA system.

		To investigate the impact of the length of the square area ${D_{\text{aera}}}$ on the performance of MA and PA systems, we report the effective rank of all algorithms for different values of ${D_{\text{aera}}}$ when the number of users ${N=80}$, as shown in \reftable{table_eRank_as_various_system_parameters}.
        As the length of the square area increases, the MA system exhibits a noticeable degradation in the effective rank, whereas the PA system demonstrates only mild fluctuations without a clear decreasing trend.
        Specifically, since the MA movement is confined to a 2D array plane at the BS, its achievable spatial DoF is fixed, causing the effective channel vectors of different users to become increasingly similar as the user region expands.
        Consequently, the inter-user channel correlation is strengthened, leading to a more concentrated eigenvalue distribution of the aggregated channel matrix and, hence, a reduction in the effective rank under the length of the square area decreasing.
        In contrast, the PA system benefits from a waveguide whose physical length can scale with the length of the square area. 
        This is because that the enlarged waveguide provides additional spatial sampling capability, preventing excessive alignment among the effective channel vectors of different users.
        Consequently, the standard deviation of the effective rank of the PA system is 0.005, compared to 0.230 for the MA system, indicating a 97.8\% improvement in terms of the stability of  the spatial DoF.

	\section{Conclusion \label{sec_conclusion}}
        This paper investigates the spatial DoF of MA and PA-enabled wireless systems that optimize antenna positions under collision-free constraints to enhance the effective rank over multi-time slots.
        The effective rank is introduced as a metric for analysis and optimization of MA and PA systems.
        For MA,  a user graph based on spatial distribution is built in the GRL scheme, and then the BS acts as an agent to use our proposed GAIQN algorithm to optimize MA positions, where its top-k action selection method ensures collision-free between multiple MAs on the 2D array plane.
        For PA,  a user graph based on waveguide region is built in the MAGRL scheme,  and then each waveguide acts as a GRL agent to use our proposed MAGAQN algorithm to optimize PA positions, where its top-k action selection method is also used to achieve collision-free between multiple PAs on the same waveguide.
        Simulation results verify that the GAIQN and MAGAQN algorithms improve the effective rank by at least 1.6\% and 1.3\% compared with benchmarks, respectively.
        Under the same number of antennas, we further examine the influence of different numbers of users and the length of the square area on MA and PA systems, and reveal the MA system supports the higher effective rank than the PA system, but the PA system offers greater stability in the achievable spatial DoF.

	\bibliographystyle{IEEEtran}
	\bibliography{ref_new20260215}

@Article{11267231,
  author   = {Xu, Yiming and Xu, Dongfang and Yu, Xianghao and Song, Shenghui and Ding, Zhiguo and Schober, Robert},
  journal  = {IEEE Trans. Wireless Commun.},
  title    = {Joint Radiation Power, Antenna Position, and Beamforming Optimization for Pinching-Antenna Systems With Motion Power Consumption},
  year     = {2026},
  pages    = {7825-7841},
  volume   = {25},
  doi      = {10.1109/TWC.2025.3633833},
  fjournal = {IEEE Transactions on Wireless Communications},
}

@Article{11142311,
  author   = {Shao, Xiaodan and Mei, Weidong and You, Changsheng and Wu, Qingqing and Zheng, Beixiong and Wang, Cheng-Xiang and Li, Junling and Zhang, Rui and Schober, Robert and Zhu, Lipeng and Zhuang, Weihua and Shen, Xuemin},
  journal  = {IEEE Commun. Surveys Tuts.},
  title    = {A Tutorial on Six-Dimensional Movable Antenna for 6G Networks: Synergizing Positionable and Rotatable Antennas},
  year     = {2026},
  pages    = {3666-3709},
  volume   = {28},
  doi      = {10.1109/COMST.2025.3602939},
  fjournal = {IEEE Communications Surveys and Tutorials},
}

@Article{11222687,
  author   = {Yang, Zheng and Wang, Ning and Sun, Yanshi and Ding, Zhiguo and Schober, Robert and Karagiannidis, George K. and Wong, Vincent W.S. and Dobre, Octavia A.},
  journal  = {IEEE Wirel. Commun.},
  title    = {Pinching Antennas: Principles, Applications and Challenges},
  year     = {2025},
  pages    = {1-10},
  doi      = {10.1109/MWC.2025.3607867},
  fjournal = {IEEE Wireless Communications},
}

@InProceedings{11143302,
  author    = {Chen, Kangjian and Qi, Chenhao and Hong, Yujing and Yuen, Chau},
  booktitle = {2025 IEEE 26th International Workshop on Signal Processing and Artificial Intelligence for Wireless Communications (SPAWC)},
  title     = {Multiuser Sum-Rate Maximization for Reconfigurable Pixel Antenna-Based Electronic Movable-Antenna Arrays},
  year      = {2025},
  pages     = {1-5},
  doi       = {10.1109/SPAWC66079.2025.11143302},
}

@Article{11048972,
  author   = {Wu, Yifei and Xu, Dongfang and Wing Kwan Ng, Derrick and Gerstacker, Wolfgang and Schober, Robert},
  journal  = {IEEE Trans. Commun.},
  title    = {Globally Optimal Movable Antenna-Enabled Multiuser Communication: Discrete Antenna Positioning, Power Consumption, and Imperfect CSI},
  year     = {2025},
  number   = {10},
  pages    = {9903-9923},
  volume   = {73},
  doi      = {10.1109/TCOMM.2025.3582717},
  fjournal = {IEEE Transactions on Communications},
}

@Article{11164786,
  author   = {Chen, Xintai and Feng, Biqian and Wu, Yongpeng and Ng, Derrick Wing Kwan and Schober, Robert},
  journal  = {IEEE Trans. Wireless Commun.},
  title    = {Two-Timescale Sum-Rate Maximization for Movable Antenna Enhanced Systems},
  year     = {2026},
  pages    = {3894-3909},
  volume   = {25},
  doi      = {10.1109/TWC.2025.3606950},
  fjournal = {IEEE Transactions on Wireless Communications},
}

@Article{11226954,
  author   = {Zhu, Lipeng and Sun, He and Ma, Wenyan and Xiao, Zhenyu and Zhang, Rui},
  journal  = {IEEE Trans. Wireless Commun.},
  title    = {Multiuser Communications Aided by Cross-Linked Movable Antenna Array: Architecture and Optimization},
  year     = {2026},
  pages    = {6729-6744},
  volume   = {25},
  doi      = {10.1109/TWC.2025.3626388},
  fjournal = {IEEE Transactions on Wireless Communications},
}

@Article{11218873,
  author   = {Chen, Guangyi and Zhang, Ruoyu and Guan, Xinrong and Hu, Guojie and Wu, Qingqing and Wu, Wen},
  journal  = {IEEE Internet Things J.},
  title    = {Energy Efficiency Maximization for Multiuser Communications With Movable Antennas: Joint Beamforming and Antenna Position Design},
  year     = {2026},
  number   = {1},
  pages    = {868-881},
  volume   = {13},
  doi      = {10.1109/JIOT.2025.3626139},
  fjournal = {IEEE Internet of Things Journal},
}

@Article{11175694,
  author   = {Zhang, Jiarui and Xu, Hao and Ouyang, Chongjun and Zou, Qiuyun and Yang, Hongwen},
  journal  = {IEEE Commun. Lett.},
  title    = {Uplink Sum-Rate Maximization for Pinching Antenna-Assisted Multiuser MISO Communications},
  year     = {2025},
  number   = {12},
  pages    = {2795-2799},
  volume   = {29},
  doi      = {10.1109/LCOMM.2025.3613565},
  fjournal = {IEEE Communications Letters},
}

@Article{11184829,
  author   = {Che, Bohan and Zhu, Lei and Cheng, Kaixin},
  journal  = {IEEE Wireless Commun. Lett.},
  title    = {Throughput Maximization in Multi-User PAS Via Joint Optimization for TDMA},
  year     = {2026},
  pages    = {415-419},
  volume   = {15},
  doi      = {10.1109/LWC.2025.3616162},
  fjournal = {IEEE Wireless Communications Letters},
}

@Article{11300296,
  author   = {Zhao, Jingjing and Song, Haowen and Mu, Xidong and Cai, Kaiquan and Zhu, Yanbo and Liu, Yuanwei},
  journal  = {IEEE Trans. Commun.},
  title    = {Pinching-Antenna Systems-Enabled Multi-User Communications: Transmission Structures and Beamforming Optimization},
  year     = {2026},
  pages    = {2316-2330},
  volume   = {74},
  doi      = {10.1109/TCOMM.2025.3643993},
  fjournal = {IEEE Transactions on Communications},
}

@Article{11095802,
  author   = {Chen, Kangjian and Qi, Chenhao and Hong, Yujing and Yuen, Chau},
  journal  = {IEEE Trans. Commun.},
  title    = {REMAA: Reconfigurable Pixel Antenna-Based Electronic Movable-Antenna Arrays for Multiuser Communications},
  year     = {2025},
  number   = {11},
  pages    = {12913-12928},
  volume   = {73},
  doi      = {10.1109/TCOMM.2025.3592593},
  fjournal = {IEEE Transactions on Communications},
}

@Article{11165763,
  author   = {Wang, Kaidi and Ding, Zhiguo and Karagiannidis, George K.},
  journal  = {IEEE Trans. Wireless Commun.},
  title    = {Antenna Activation and Resource Allocation in Multi-Waveguide Pinching-Antenna Systems},
  year     = {2026},
  pages    = {4070-4082},
  volume   = {25},
  doi      = {10.1109/TWC.2025.3608068},
  fjournal = {IEEE Transactions on Wireless Communications},
}

@InProceedings{10217398,
  author    = {Ma, Jinyan and Li, Ruifeng and Li, Da and Zhang, Ling and Li, Erping},
  booktitle = {2023 Joint Asia-Pacific International Symposium on Electromagnetic Compatibility and International Conference on ElectroMagnetic Interference \& Compatibility (APEMC/INCEMIC)},
  title     = {Position-Based Optimization of the Electromagnetic Channel Effective Rank for MIMO Systems},
  year      = {2023},
  pages     = {1-4},
  doi       = {10.1109/APEMC57782.2023.10217398},
}

@Article{10314137,
  author   = {Meng, Shengguo and Tang, Wankai and Chen, Weicong and Lan, Jifeng and Zhou, Qun Yan and Han, Yu and Li, Xiao and Jin, Shi},
  journal  = {IEEE Wireless Commun. Lett.},
  title    = {Rank Optimization for MIMO Channel With RIS: Simulation and Measurement},
  year     = {2024},
  number   = {2},
  pages    = {437-441},
  volume   = {13},
  doi      = {10.1109/LWC.2023.3331489},
  fjournal = {IEEE Wireless Communications Letters},
}

@Article{11225901,
  author   = {Sun, Qiang and Cao, Ye and Li, Zejun and Chen, Xiaomin and Li, Dong and Zhang, Jiayi},
  journal  = {IEEE Trans. Commun.},
  title    = {Effective Rank Maximization for Active RIS-Assisted MIMO Systems},
  year     = {2026},
  pages    = {536-550},
  volume   = {74},
  doi      = {10.1109/TCOMM.2025.3628735},
  fjournal = {IEEE Transactions on Communications},
}

@InProceedings{10304448,
  author    = {Zhu, Lipeng and Ma, Wenyan and Ning, Boyu and Zhang, Rui},
  booktitle = {2023 IEEE 24th International Workshop on Signal Processing Advances in Wireless Communications (SPAWC)},
  title     = {Multiuser Communication Aided by Movable Antenna},
  year      = {2023},
  pages     = {531-535},
  doi       = {10.1109/SPAWC53906.2023.10304448},
}

@Article{10416896,
  author   = {Qin, Haoran and Chen, Wen and Li, Zhendong and Wu, Qingqing and Cheng, Nan and Chen, Fangjiong},
  journal  = {IEEE Wireless Commun. Lett.},
  title    = {Antenna Positioning and Beamforming Design for Fluid Antenna-Assisted Multi-User Downlink Communications},
  year     = {2024},
  number   = {4},
  pages    = {1073-1077},
  volume   = {13},
  doi      = {10.1109/LWC.2024.3360117},
  fjournal = {IEEE Wireless Communications Letters},
}

@Article{11223640,
  author   = {Sun, Mingjun and Ouyang, Chongjun and Wu, Shaochuan and Liu, Yuanwei},
  journal  = {IEEE Trans. Wireless Commun.},
  title    = {Multiuser Beamforming for Pinching-Antenna Systems: An Element-Wise Optimization Framework},
  year     = {2026},
  pages    = {6538-6552},
  volume   = {25},
  doi      = {10.1109/TWC.2025.3625377},
  fjournal = {IEEE Transactions on Wireless Communications},
}

@Article{papanikolaou2025physical,
  author  = {Papanikolaou, Pigi P and Bozanis, Dimitrios and Tegos, Sotiris A and Diamantoulakis, Panagiotis D and Sarigiannidis, Panagiotis and Karagiannidis, George K},
  journal = {arXiv preprint arXiv:2511.23079},
  title   = {Physical Layer Security with Artificial Noise in MIMO Pinching-Antenna Systems},
  year    = {2025},
}

@InProceedings{roy2007effective,
  author       = {Roy, Olivier and Vetterli, Martin},
  booktitle    = {2007 15th European signal processing conference},
  title        = {The effective rank: A measure of effective dimensionality},
  year         = {2007},
  organization = {IEEE},
  pages        = {606--610},
}

@Article{11020999,
  author   = {Le Hung, Hoang and Huy, Nguyen Hoang and Luong, Nguyen Cong and Pham, Quoc-Viet and Niyato, Dusit and Hoa, Nguyen Tien},
  journal  = {IEEE Trans. Veh. Technol.},
  title    = {Beamforming Design for Physical Security in Movable Antenna-Aided ISAC Systems: A Reinforcement Learning Approach},
  year     = {2025},
  number   = {11},
  pages    = {18163-18167},
  volume   = {74},
  doi      = {10.1109/TVT.2025.3575653},
  fjournal = {IEEE Transactions on Vehicular Technology},
}

@Article{9930883,
  author   = {Feres, Carlos and Ding, Zhi},
  journal  = {IEEE Trans. Wireless Commun.},
  title    = {An Unsupervised Learning Paradigm for User Scheduling in Large Scale Multi-Antenna Systems},
  year     = {2023},
  number   = {5},
  pages    = {2932-2945},
  volume   = {22},
  doi      = {10.1109/TWC.2022.3215471},
  fjournal = {IEEE Transactions on Wireless Communications},
}

@Article{velivckovic2017graph,
  author  = {Veli{\v{c}}kovi{\'c}, Petar and Cucurull, Guillem and Casanova, Arantxa and Romero, Adriana and Lio, Pietro and Bengio, Yoshua},
  journal = {arXiv preprint arXiv:1710.10903},
  title   = {Graph attention networks},
  year    = {2017},
}

@InProceedings{dabney2018implicit,
  author       = {Dabney, Will and Ostrovski, Georg and Silver, David and Munos, R{\'e}mi},
  booktitle    = {International conference on machine learning},
  title        = {Implicit quantile networks for distributional reinforcement learning},
  year         = {2018},
  organization = {PMLR},
  pages        = {1096--1105},
}

@Article{10540320,
  author   = {Wu, Changmao and Xu, Zhengwei and He, Xiaoming and Lou, Qi and Xia, Yuanyuan and Huang, Shuman},
  journal  = {IEEE Trans. Parallel Distrib. Syst.},
  title    = {Proactive Caching With Distributed Deep Reinforcement Learning in 6G Cloud-Edge Collaboration Computing},
  year     = {2024},
  number   = {8},
  pages    = {1387-1399},
  volume   = {35},
  doi      = {10.1109/TPDS.2024.3406027},
  fjournal = {IEEE Transactions on Parallel and Distributed Systems},
}

@Article{11320569,
  author   = {Yu, Jiaming and Liang, Le and Ye, Hao and Jin, Shi},
  journal  = {IEEE Trans. Mob. Comput.},
  title    = {Hierarchical Multi-Agent Reinforcement Learning-based Coordinated Spatial Reuse for Next Generation WLANs},
  year     = {2025},
  pages    = {1-16},
  doi      = {10.1109/TMC.2025.3649899},
  fjournal = {IEEE Transactions on Mobile Computing},
}

@Article{9839534,
  author   = {Chen, Wanghu and Zhai, Chenhan and Wang, Xin and Li, Jing and Lv, Pengbo and Liu, Chen},
  journal  = {IEEE Trans. Ind. Inf.},
  title    = {GCN- and GRU-Based Intelligent Model for Temperature Prediction of Local Heating Surfaces},
  year     = {2023},
  number   = {4},
  pages    = {5517-5529},
  volume   = {19},
  doi      = {10.1109/TII.2022.3193414},
  fjournal = {IEEE Transactions on Industrial Informatics},
}

@Article{chung2014empirical,
  author  = {Chung, Junyoung and Gulcehre, Caglar and Cho, KyungHyun and Bengio, Yoshua},
  journal = {arXiv preprint arXiv:1412.3555},
  title   = {Empirical evaluation of gated recurrent neural networks on sequence modeling},
  year    = {2014},
}

@Article{10741192,
  author   = {Xiao, Zhenyu and Pi, Xiangyu and Zhu, Lipeng and Xia, Xiang-Gen and Zhang, Rui},
  journal  = {IEEE Trans. Wireless Commun.},
  title    = {Multiuser Communications With Movable-Antenna Base Station: Joint Antenna Positioning, Receive Combining, and Power Control},
  year     = {2024},
  number   = {12},
  pages    = {19744-19759},
  volume   = {23},
  doi      = {10.1109/TWC.2024.3486320},
  fjournal = {IEEE Transactions on Wireless Communications},
}

@TechReport{3gpp.38.901,
  author      = {3GPP},
  institution = {{3rd Generation Partnership Project (3GPP)}},
  title       = {{Study on channel model for frequencies from 0.5 to 100 GHz (Release 19)}},
  year        = {2025},
  note        = {Version 19.2.0},
  number      = {38.901},
  type        = {Technical Specification (TS)},
  url         = {http://www.3gpp.org/
	ftp//Specs/archive/38_series/38.901/},
}

@Article{yu2023enhancing,
  author  = {Yu, Peihong and Lee, Bhoram and Raghavan, Aswin and Samarasekara, Supun and Tokekar, Pratap and Hare, James Zachary},
  journal = {arXiv preprint arXiv:2311.04740},
  title   = {Enhancing Multi-Agent Coordination through Common Operating Picture Integration},
  year    = {2023},
}

@Article{yang2020qatten,
  author  = {Yang, Yaodong and Hao, Jianye and Liao, Ben and Shao, Kun and Chen, Guangyong and Liu, Wulong and Tang, Hongyao},
  journal = {arXiv preprint arXiv:2002.03939},
  title   = {Qatten: A general framework for cooperative multiagent reinforcement learning},
  year    = {2020},
}

@Article{yu2022surprising,
  author  = {Yu, Chao and Velu, Akash and Vinitsky, Eugene and Gao, Jiaxuan and Wang, Yu and Bayen, Alexandre and Wu, Yi},
  journal = {Advances in neural information processing systems},
  title   = {The surprising effectiveness of ppo in cooperative multi-agent games},
  year    = {2022},
  pages   = {24611--24624},
  volume  = {35},
}

@Article{de2020independent,
  author  = {De Witt, Christian Schroeder and Gupta, Tarun and Makoviichuk, Denys and Makoviychuk, Viktor and Torr, Philip HS and Sun, Mingfei and Whiteson, Shimon},
  journal = {arXiv preprint arXiv:2011.09533},
  title   = {Is independent learning all you need in the starcraft multi-agent challenge?},
  year    = {2020},
}
\end{document}